**The Influence of Dimensionality on the Charge Density Wave Transition and Its Application on Mid-infrared Photodetection**

*Jialin Li, Hua Bai, Yupeng Li, Junjian Mi, Qiang Chen, Wei Tang, Huanfeng Zhu, Xinyi Fan, Yunhao Lu*, Zhuan Xu*, Linjun Li**

Jialin Li, Qiang Chen, Wei Tang, Huanfeng Zhu, Xinyi Fan

State Key Laboratory of Modern Optical Instrumentation, College of Optical Science and Engineering, Zhejiang University, Hangzhou 310027, China.

Hua Bai, Yupeng Li, Junjian Mi, Yunhao Lu, Zhuan Xu, Linjun Li

Zhejiang Province Key Laboratory of Quantum Technology and Device, Department of physics, Zhejiang University, Hangzhou 310027, China.

Yunhao Lu, Zhuan Xu

State Key Laboratory of Silicon Materials, Zhejiang University, Hangzhou 310027, China.

Huanfeng Zhu, Linjun Li

Intelligent Optics & Photonics Research Center, Jiaxing Research Institute Zhejiang University, Jiaxing 314000, China.

Huanfeng Zhu, Linjun Li

Jiaxing Key Laboratory of Photonic Sensing & Intelligent Imaging, Jiaxing Institute Zhejiang University, Jiaxing, 314000 China.

*Email: luyh@zju.edu.cn; zhuan@zju.edu.cn; lilinjun@zju.edu.cn

**Abstract**

Two-dimensional charge density wave (CDW) materials received much attention for high responsivity and broadband photodetection in recent years, due to their collective electron transport and narrow bandgap. However, the high dark current density problem hinders their real application. Here we report a sharp CDW transition in quasi-1D $(TaSe_4)_2I$, and apply it for broadband photodetection. Especially at mid-infrared region, the device shows both high photo responsivity of 1.18 $\times 10^3$ A/W and large light on/off ratio of 80, which is superior than 2D CDW $TaS_2$ and most reported low-dimensional materials. The fact for such high performance lies on two aspects. One is the much lower dark current density resulted from the pseudo gap associated with 1D Luttinger liquid state, which is supported by finite size scaling of nonlinear *I~V* at variable temperatures and occurrence of 1D structural phase transition consolidated by In-situ Raman spectroscopy. The other is the high photocurrent associated with the "Fröhlich superconductivity" state, manifested by an ultrasensitive switching, which can be only accessible in 1D CDW materials, in agreement with our density functional theory calculation. Our work thus reveals the pivotal role of dimensionality in CDW phase transition, and paves a way for implementing highly sensitive broadband photodetector.

Keywords: Dimensionality; Charge density wave; Mid-infrared photodetection; Quasi 1D; $(TaSe_4)_2I$

## 1. Introduction

Broadband photodetectors, especially the mid-infrared ones, have been the heart of various fields such as biomedicine, thermo imaging, and remote sensing. The commercialized photodetectors for mid-infrared (MIR) range such as InGaAs, HgCdTe and InSb often suffer from strict growth procedure, working at liquid nitrogen temperature and hence high cost[1]. Avalanche photodetectors or photodiodes (APD) enabled by impact ionization mechanism, can achieve large carrier multiplication with high gain, providing a strategy for high performace MIR photodetection, which has been achieved in HgCdTe-based APDs[2], multi-quantum well (MQW) heterostructures[3] and recently in InSe/BP heterostructure[4]. However, either the low working temperature or the ordinary photo responsivity (below 10 A/W) calls for seeking other candidate materials.

CDW materials are famous for their collective electronic excitations in moderate temperature, attracting much attention recently because of their potential applications in electronics[5], memory[6] and photodetection devices[7]. They usually harbor a low dimensional structure (quasi 1D or 2D) since the stronger phonon softening in low dimensional plays the key role in electron phonon coupling which responds to occurance of CDW[8]. When CDW materials cooled below a critical temperature $T_{CDW}$, the crystalline lattice is usually distorted under the electron-phonon interaction while the electron density is periodically modulated in real space and mildly gapped in momentum space. Manipulating this collective condensation by electric field, photoexcitation is accompanied by a large variation of conductance, usually several orders of magnitudes, which enables a high responsivity in a broadband region. For instance, Dong Wu[7a] reported broadband photodetection from visiable to terahertz by exploiting 1T-$TaS_2$ device with high responsivity of ~1 A/W, attracting the attention on the photodetection exploiting CDW materials[7b-d]. Nevertheless, those photodetectors based on 2D CDW materials experience high dark current density and low light on/off ratio even at liquid nitrogen temperature, which seriously deteriorate their performance. Futhermore, due to the random distribution of CDW domains in $TaS_2$[9], the device yield is quite low, which is very unstable working as a photodetector. While in recent years the 2D CDW materials have been intensively investigated for photodetection, the 1D CDW materials are seldomly explored. According to the literatures[10], the photoresponse of 1D CDW materials was investigated several decades ago, with the limitation of light wavelength below 1 $\mu$m, and was not in purpose survey for photodetetion. With certain CDW domains aligned along the chains, 1D CDW materials own steeper and larger amplitude in resistance switch[10a, 11] than the 2D CDW materials, it is thereofore expectable to reach higher photoresponsivity by exploiting 1D CDW material for photodetection. Furthermore, it was reported for 1D CDW materials, pseudogap related semiconducting behavior instead of metallic ground state exists above CDW transition temperature[12], which hints for a possible lower dark current in the photodetection based on 1D CDW transition. Recently, low-noise current level characteristics at dark state were observed in quasi-1D (TaSe4)2I[13], which would be benefit for obtaining high signal-to-noise ratio in photodetection application. Furthermore, high performance photodetection up to 28 A/W at room temperature has been reported in (TaSe4)2I nanowire[14], but still limited in visiable to near-infrared region. These motivate us to investigate 1D CDW materials, i.e. (TaSe4)2I, for MIR photodetection to realize both high photoresponsivty and high light on/off ratio (Figure 1a)

Herein, we investigated the photodetection of quasi 1D CDW (TaSe4)2I in broadband, especially focused on MIR region. We reported the observation of avalanche-like phenomenon in 1D (TaSe4)2I, and achieved both high photo responsivity of 1.18×103 A/W and large light on/off ratio of 80 at MIR region. Compared to 2D CDW TaS2, as expectation, three orders of magnitude lower dark current density and two orders of magnitude higher photocurrent were achieved. To the best of our knowledge, our work is the first one to investigate the dimensionality influence on the charge density wave transition for mid-infrared photodetection and achieved such high MIR photodetection performance in quasi 1D CDW material, thus paves the way for exploiting them on advanced optoelectronic application.

## 2. Results and Discussion

The conventional cell of (TaSe4)2I contains two TaSe4 chains aligned along the c-axis and four iodine atoms separating the chains, which belong to a tetragonal structure at room temperature[15]. Each chain contains a total of 4 Ta atoms and 16 Se atoms (Figure 1b, left). Needle-like single crystals of (TaSe4)2I were obtained via a chemical vapour transport method. The linear |TaSe4| chain-like single crystal structure was characterized through high-resolution transmission electron microscope (HRTEM) image (Figure S1a,b). The energy-dispersive X-ray spectroscopy reveals a Ta:Se:I ratio of 2.1:7.8:1 (Figure S1c) and the X-ray diffraction (XRD) spectrum also confirms the structure of (TaSe4)2I (Figure S2a). Broadband absorption from ultraviolet to long infrared region were revealed by the measured transmittance spectrum (Figure S2b). Raman scattering measurement on $(TaSe_4)_2I$ single crystal in Figure S2c shows that the obvious 6 peaks (68.97 $cm^{-1}$, 101.15 $cm^{-1}$, 146.36 $cm^{-1}$, 158.76 $cm^{-1}$, 182.72 $cm^{-1}$, 271.45 $cm^{-1}$) corresponding to three vibration modes. $A_1$, $B_2$, and E , agree with the previously reported result[16]. The anisotropic quasi-1D crystal structure was also confirmed by polarized Raman measurement (Fig.S2d). Figure 1c shows the four-probe temperature-dependent resistivity and carrier density of $(TaSe_4)_2I$, as temperature decreased to the transition temperature of $T_c$ = 261 K, $(TaSe_4)_2I$ undergoes a Peierls-like transition, forming an incommensurate CDW phase, which is accompanied by opening a gap of ~266 meV, with sharp carrier density variation of 4 orders at low temperature. Since the CDW instability corresponds to the Ta-tetramerization modes[17], a long-long-short-short (LLSS) Ta-tetramerization CDW pattern is expected (Figure 1b, right). Therefore it naturally provides a platform to manipulate the phase transition between CDW state and metal state via applying electric field or photoexcitation.

### 2.1. Extremely-Sharp-Current-Jump (ESCJ) Driven by Electric Field

To investigate the low temperature optoelectronic properties of $(TaSe_4)_2I$, we selected two samples with different sample quality (due to the different growth condition as describled in Experimental section) as representatives, which were refered as sample1 (unoptimzed) and sample2 (defect-optimzed). The representative two-probe I-V curves of sample2 measured at different temperature are plotted in Figure 1d, both voltage sweeping forward and backward were performed in voltage driving mode. As the temperature cools below $T_c$, we observe nonlinear and hysteretic *I-V* behavior when the applied electric field exceeds a voltage threshold. At lower temperature (i.e. 120 K), we observed a ESCJ phenomenon when voltage exceeds a threshold. The whole *I-V* curve can be divided into four different intervals: a linear section at low voltage bias is a high resistive state, a nonlinear section when exceeds a voltage bias threshold $V_T$, followed by a huge current jump-up to a low resistive state when just exceeds a threshold $V_T^*$ and a linear section of a metal state thereafter. Furthermore, we still observe the ESCJ phenomenon though minimizing the voltage interval to 10 μV which is the precision limit of our sourcemeter, as Figure S3e shows. The voltage threshold and hysteretic window becomes larger as the temperature decreases (Figure 1d and Figure S3). We could get higher high-low (on-off) resistance ratio of the transition by decreasing temperature, though it will cost larger power consumption. Such nonlinear electrical transport properties can be explained by the two-fluid model. Acrossing CDW transition temperature, there are two types of conductive mechanism involved, which named normal carrier (defects and thermal activated) and CDW excited electrons. CDW must overcome the pinning potential caused by impurities and defects to produce global sliding motion and finally free sliding[18] . When applying a low electric field on the sample, CDW is pinned by pinning potential, only normal carriers involve in conduction, in accordance with ohm's law, which is called CDW pinning region as insert of Fig1d shows. While applied bias voltage reaches a threshold value $V_T$, CDW electrons can overcome pinning potential, gradually slide and excited electrons participate in conduction, thus a nonlinear *I-V* appears. At a higher voltage up to $V_T$*, the sharp transition happens, which is called CDW free sliding region. In 1D CDW materials, this ESCJ is interpreted as possible “Fröhlich superconductivity” state[8, 18a], where CDW electrons can move without damping and lead to an infinite dynamic conductance. After ESCJ, a linear part called the metal state region follows, similar as $K_{0.3}MoO_3$[18a]. Previously, it has been revealed for an electric field driven *I-V* switching behavior at 180 K, characterizing a CDW sliding motion in the $(TaSe_4)_2I$ system[19]. However, their result did not show the ESCJ phenomena, because the screening effect of normal carriers could not be neglected at relatively high temperature.

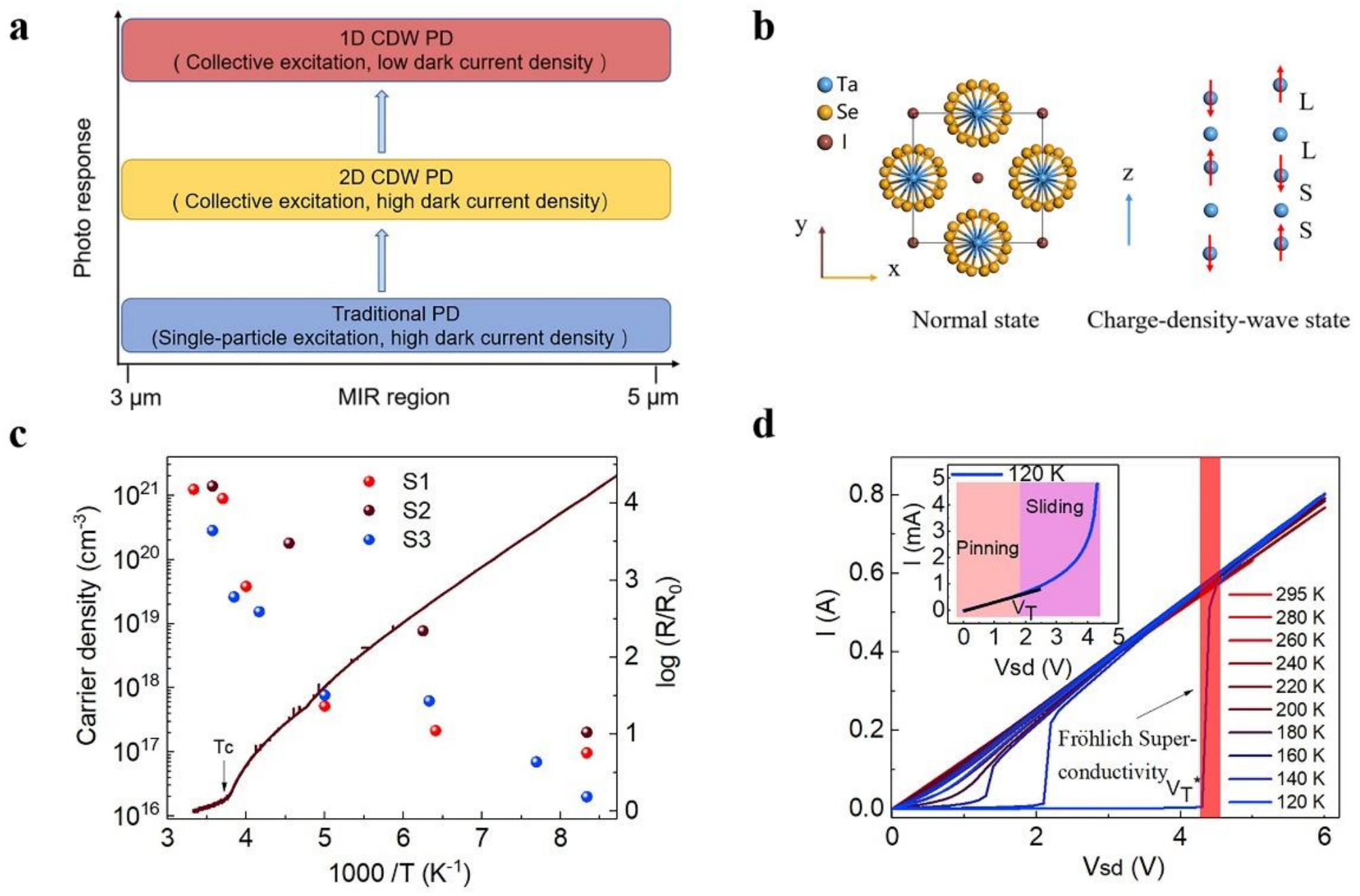

**Figure 1.** The crystal structure and electric transport measurement of $(TaSe_4)_2I$. a) Comparison of MIR photodetectors (PD) based on single material. Only intrinsic photo responsivity of single material is considered here (without any treatment like ion implantation, adding plasmonic structure or ferroelectric polymers). b) Schematic images of crystal structure of $(TaSe_4)_2I$. c) Temperature-dependent resistivity and carrier density of $(TaSe_4)_2I$. d) The representative *I-V* curves measured at different temperatures under dark state with voltage sweeping forward and backward. The insert figure shows that the linear and nonlinear behavior under bias voltage at 120 K.

**2.2. Superior Performance of Broadband Photodetection Based on ESCJ:**

Figure 3a depicts a typical photocurrent response of $(TaSe_4)_2I$ (sample1) under laser irradiation of 532 nm at 120 K. The threshold voltage $V_T$* of ESCJ in the *I-V* curves of $(TaSe_4)_2I$ is obviously suppressed by laser illumination, which demonstrates that light can also promote the CDW to normal metal state transition, as already reported in 2D $TaS_2$ previously[7a]. Such photo-reshaped CDW transition becomes more obvious when temperature decreases (Figure S6). To characterize the photo responsivity of the device, a fixed bias voltage below $V_T$* mode is employed. Upon increasing incident light intensity, a bias voltage dependent ESCJ like transition is triggered at a critical intensity $P_c$ in I-P curves, as shown in insert of Figure 3b. The photo responsivity is calculated by $R = I_{pc}/P$, where $I_{pc}$ is the net photocurrent, which is defined as the photocurrent subtracted by the dark current. Maximum photo responsivity can be derived at this sharp transition point. For instance, huge photo responsivity of 405 A/W under 3.5 V bias voltage is achieved for 532 nm laser illumination. The photo responsivity can be further increased by decreasing the work temperature or adjusting the bias voltage closer to $V_T^*$. Due to the hysteretic behavior of the phase transition, the resistance of the $(TaSe_4)_2I$ did not recover to the initial value above $P_c$ (Figure S8). Similar result for 1064 nm laser illumination is also shown in Figure S9. For practical photodetector application, the synchronized pulse voltage working mode is necessary to reset the metal state to insulating state, which has been used in previous work of $TaS_2$[7b, 7c], and also commonly used in single photon avalanche diodes (SPAD) sensors[20]. In this mode, photocurrent could be defined as $I_{pc}=I_{pulse_light}-I_{pulse_dark}$, and the light on/off ratio is defined as $I_{pulse_light}/I_{pulse_dark}$. We found that the photo-reshaped CDW transition of bulk $(TaSe_4)_2I$ crystal can be extended to MIR wavelength range, the photo responsivity reaches ~ 31.67 A/W and light on/off ratio is about 30 at $\lambda = 4.73$ μm, which is about one order of magnitude larger than that of other CDW material reported, such as $TaS_2$[7a]. The broadband photo responsivity from visible to MIR range with bias voltage of 3.5 V is summarized in Figure 3b. The photo responsivity and light on/off ratio can be further optimized by improving the sample quality, for example, we achieve both high photo responsivity of 1.18 $\times 10^3$ A/W and large light on/off ratio of 80 at $\lambda = 4.64$ μm in sample2 (detailed photoresponse seen in Fig.5a,b), which is superior than 2D CDW TaS2 and most reported low-dimensional materials at MIR region. Also, as expectation, low dark current density of 1.846 A/cm2 was achieved, which is three orders of magnitude lower than the bulk sample of TaS2 reported[7a](~ 5797 A/cm2).

Figure S10a also shows the typical mid-infrared photoresponse realized in a $(TaSe_4)_2I$ nanoribbon device with width of 400 nm and the thickness of 48 nm. The dark current density is ~1 μA/μm$^2$ under bias voltage of 2 V, which is four orders lower than 2D $TaS_2$[5b]. The obtained photo responsivity is 3.62 A/W (Sample3). It can reach 61.62 A/W by further optimizing the device geometry (Figure S11, sample4). At present, the nano-thick devices could breakdown in high electric field (ESCJ) region at 120 K due to the Joule heating result from the high current density, which hinders us to achieve better stable performance than the bulk devices as expected. However, the problem can be solved by setting a current limitation to the device or decreasing the working temperature. Figure 2c and 2d shows the ESCJ phenomena realized in a nanoplate (Sample5) worked at 120 K and a nanoribbon device (Sample6) worked at 17 K respectively. The maximum photo responsivity can be reached as 1.1 × 106 A/W (Vsd=3 V) and 3.4 × 105 A/W (Vsd=10 V) respectively under λ=4640 nm excitation. Though the working temperature is relatively low, the photo responsivity is almost the recorded value in reported low-dimensional materials at MIR region (Table S1 and S2).

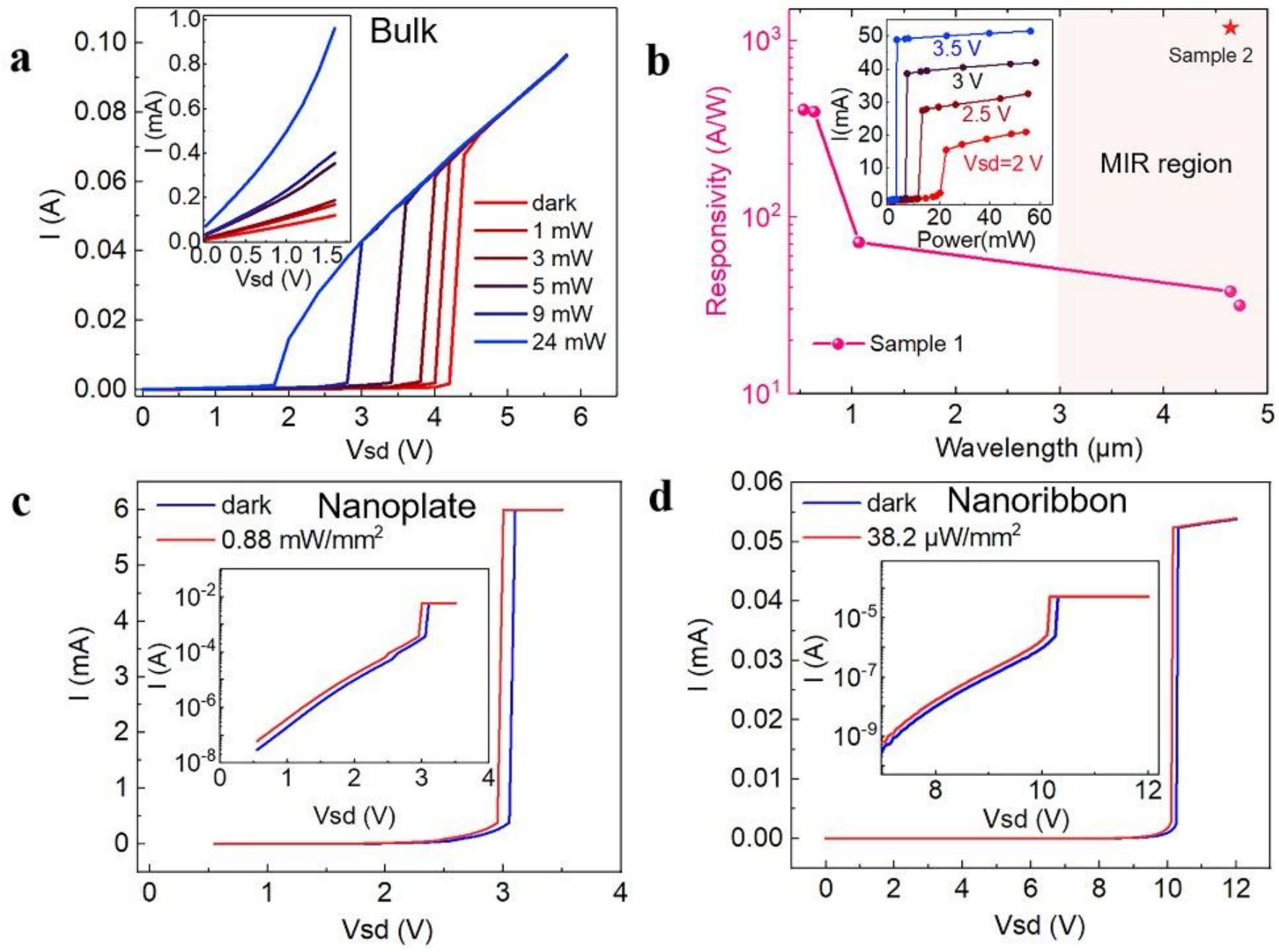

**Figure 2.** Photo electric transport of $(TaSe_4)_2I$ devices. a) Typical photocurrent response (Sample1) under a laser illumination of λ=532 nm at 120 K. The light intensity is tuned from 1 mW to 24 mW. The diameter of spot size is about 1 mm, the lateral channel size is 1 mm × 150 μm. b) The broadband photo responsivity from visible to mid-infrared range with a fixed

voltage of 3.5 V near $V_T$. Inset: The current response as the function of light power was investigated at illumination of $\lambda$=532 nm at 120 K for different applied voltages. The pentagon (responsivity) labled data is measured from another optimized high-quality sample with less defects (Sample2). c,d) ESCJ phenomena realized in a nanoplate worked at 120 K and a nanoribbon device worked at 17 K respectively (Insert figure is the log scale of *I-V* curves).

2.3. Quasi 1D Structural/electronic Phase Transition Revealed by In-situ Raman Spectroscopy and Density Functional Theory (DFT) Calculation

Since only in 1D materials, the phonon frequency at the Kohn anomaly varies sharply and can reach zero, which enables the stability of Peierls gap and ocurrance of possible Fröhlich superconductivity (FS)[8]. Such dimensionality dependence plays the key role in the CDW electrical/photo response, which results in the better photodetection performance of 1D CDW materials against 2D ones. We carried out in-situ Raman spectroscopy measurement at low temperatures. Nanoplate of $(TaSe_4)_2I$ with thickness around 500 nm was chosen to get optimized Raman signal, because the CDW peak (short for CDW related Raman peak) is weak in thick samples[16] at 120 K. In contrast to 2D CDW material, the quasi-1D $(TaSe_4)_2I$ is highly anisotropic, the CDW peak which belongs to the totally symmetric mode, was observed only in certain polarization configuration[16]. As Figure 2a shows, CDW peak near 80 $cm^{-1}$ appears at 120 K without voltage bias as the temperature decreases, consistent with the theoretical calculated Raman spectrum (Figure S4). As the voltage bias increases from 0 to 8V, the CDW peak near 80 $cm^{-1}$ is suppressed, as inset of Figure 2b shows, the peak intensity of the 80 $cm^{-1}$ decreases strikingly, which is a signature of phase transition to non-CDW state. While the voltage bias sweeps back to 0V, the CDW peak restores to original strength. Compared to 2D TaS2[9], such CDW mode related phonon soften and restore in 1D (TaSe4)2I appears much sharper when responses to external stimuli. In the following, we also distinguish such difference between 1D and 2D CDW materials confirmed by DFT calculation, when their coupled electron phonon system response to the external applied electrical potential.

The DFT result displays that the band structure is very sensitive to the variation of the Fermi-Dirac distribution around the Fermi surface (parameter σ) (Figure S5). When applied voltage bias reaches $V_T$*, the change of Fermi-Dirac distribution reaches a critical point, the Fermi vector nesting is destroyed by the reshaped Fermi surfaces, the Γ - Z gap suddenly closes. The crystal lattice needs to restore its ground state to compensate the elevated electron energy. As the upper panel of Figure2d shows, the structural transition of $(TaSe_4)_2I$ from space group No. 22 (CDW phase) to No. 97 (the semimetal phase) is manifested by the sharp decrease of the difference between L and S (L is short for long Ta-Ta distance, S is short for short Ta-Ta distance, as Figure 2c shows). Therefore, it is expectable for 1D (TaSe4)2I to approach "Fröhlich superconductivity" state (restored crystal lattice and free sliding CDW) under photo/electric driven, with ultrahigh photoresponsivity. In contrast, in 2D CDW material, the Ta-S distance of $TaS_2$ has much more robust dependence on the Fermi-Dirac distribution as shown in Figure 2d lower panel.

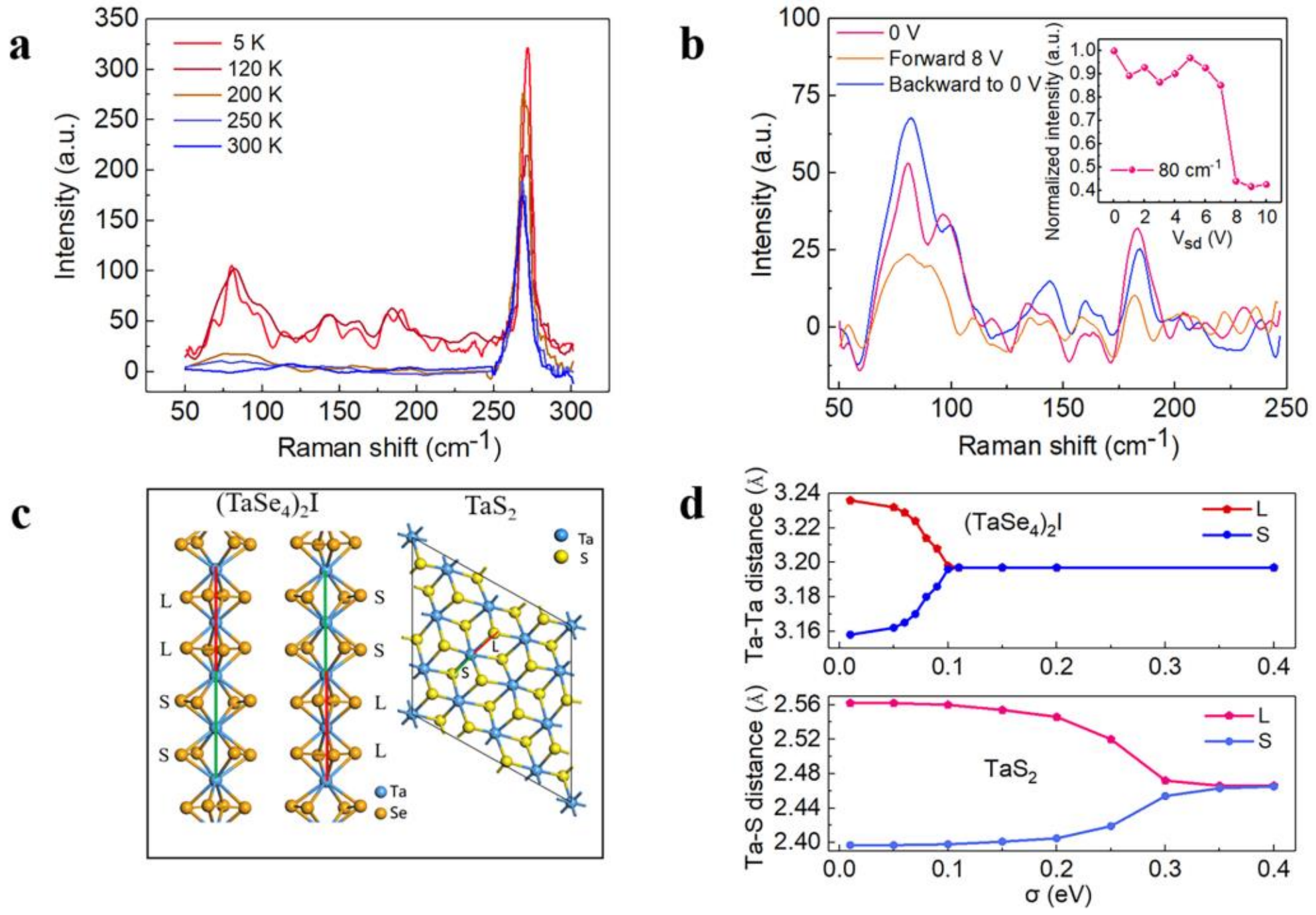

**Figure 3.** In-situ Raman measurement and DFT calculation. a) The in-situ Raman spectrum under different temperature. b) The in-situ Raman spectrum under different voltage at 120

K. Inset: The evolution of CDW Raman intensity with different applied voltage at 120 K. c) Schematic images of crystal structure of $(TaSe_4)_2I$ (To clearly show the Ta spacing in the CDW state of $(TaSe_4)_2I$, only two $TaSe_4$ chains are drawn and the rest part is omitted) and 1T-$TaS_2$. d) Ta-Ta distance vary with the value of parameter σ in $(TaSe_4)_2I$ (upper) and Ta-S distance vary in 1T-$TaS_2$ (lower).

2.4. Presence of Pseudo Gap Associated with Luttinger liquid (LL) state Above TC

As lower dark current density is indeed revealed in our 1D (TaSe4)2I, the presence of pseudo gap needs to be confirmed to corroborate such superior performance. Previous studies reported clear non-Fermi liquid lineshapes behavior in quasi 1D Peierls systems like $K_{0.3}MoO_3$, TTF–TCNQ, and $(TaSe_4)_2I$ through temperature dependent angle-resolved photoemission spectroscopy (ARPES) measurements[12], which claimed a LL phase in the normal state with a pseudogap. For consolidating the LL behavior, we investigated the temperature dependent electrical transport in exfoliated $(TaSe_4)_2I$ nano-thick device. Above $T_{CDW}$, we observe a power law scaling behavior of the transport measurement results, which can be precisely described by the LL model. Figure 4a shows the typical *I-V* curves of the device under the temperature range from 295 to 260 K, where they all show a nonlinear behavior and are less conductive with decreasing temperature. The conductance $G = I_{sd}/V_{sd}$ shows a power–law behavior with relation as $G \sim T^{\alpha}$, obtaining a fitting exponent $\alpha = 3.47$ at $V_{sd} = 0.13$ V, as Fig 4b shows. Similarly, the *I-V* curve fitting gives out an exponent $\beta = 1.49$ at 260 K, as shown in Figure 4c. Another vigorous evidence for LL behavior is the universal scaling curve of all *I-V* curves at different temperatures, which can be well depicted as the equation[21]:

$$I_{ds} = I_0 T^{1+\alpha} \sinh(\gamma e V_{sd} / k_B T) \left| \Gamma\left(\frac{1+\beta}{2} + i\gamma \frac{eV_{sd}}{\pi k_B T}\right) \right|^2 , \qquad (1)$$

Where $I_0$ is a fitting factor, $k_B$ is the Boltzmann constant, α and β are the tunneling exponents as mentioned above, and $\gamma^{-1}$ denotes the number of tunnel barriers in the transport path of Luttinger liquid. As shown in Figure 4d, our result can be well scaled and fitted with Eq. (1), with exponents of $\alpha = 3.47$, $\beta = 1.49$, and $\gamma^{-1} = 16.67$. Based on the extracted α value, the electron interaction parameter $g = \frac{1}{4\alpha + 1}$ can be calculated to be 0.067. It demonstrates that the Luttinger parameter $g \ll 1$, indicating a strong electron–electron repulsive interactions along 1D charge transport path[21c]. More experiment data with different sample width for consolidating LL behavior can be found in Figure S13 and Figure S14. The fitting exponent β for the LL behavior increases as the decreasing sample width as shown in insert figure of Figure 4c, indicating that less electron hopping possibility between along c-axis lattice chains of $(TaSe_4)_2I$, which is also a signature of LL behavior in 1D system. The results agree with another nanoscale samples of quasi-1D CDW material ($NbSe_3$) with the similar exponent and dependence on sample transverse dimension[22].

Confirmed the fact that the normal state of $(TaSe_4)_2I$ is LL state, it is apparent that the ocurrance of several orders of magnitudes higher carrier density variation accompanying the CDW transition compared to that of quasi-2D CDW materials. Since the presence of a pseudogap in LL state above CDW transition temperature[23], as shown in Figure. 4e, compared to 2D Fermi liquid case, such pseudogap further reduces the density of states near the CDW gap, leading to reduced dark current. There are also reports on the observation of pseudogap in some 2D CDW materials, such as, $TaSe_2$[24] and cuprates[25]. Nevertheless, such pseudogap in 2D material is usually anistropic, therefore, the dark current is still normally high since the transport current takes the contribution of all directions. While in 1D case, the pseudogap takes robust effect on suppressing the dark current density.

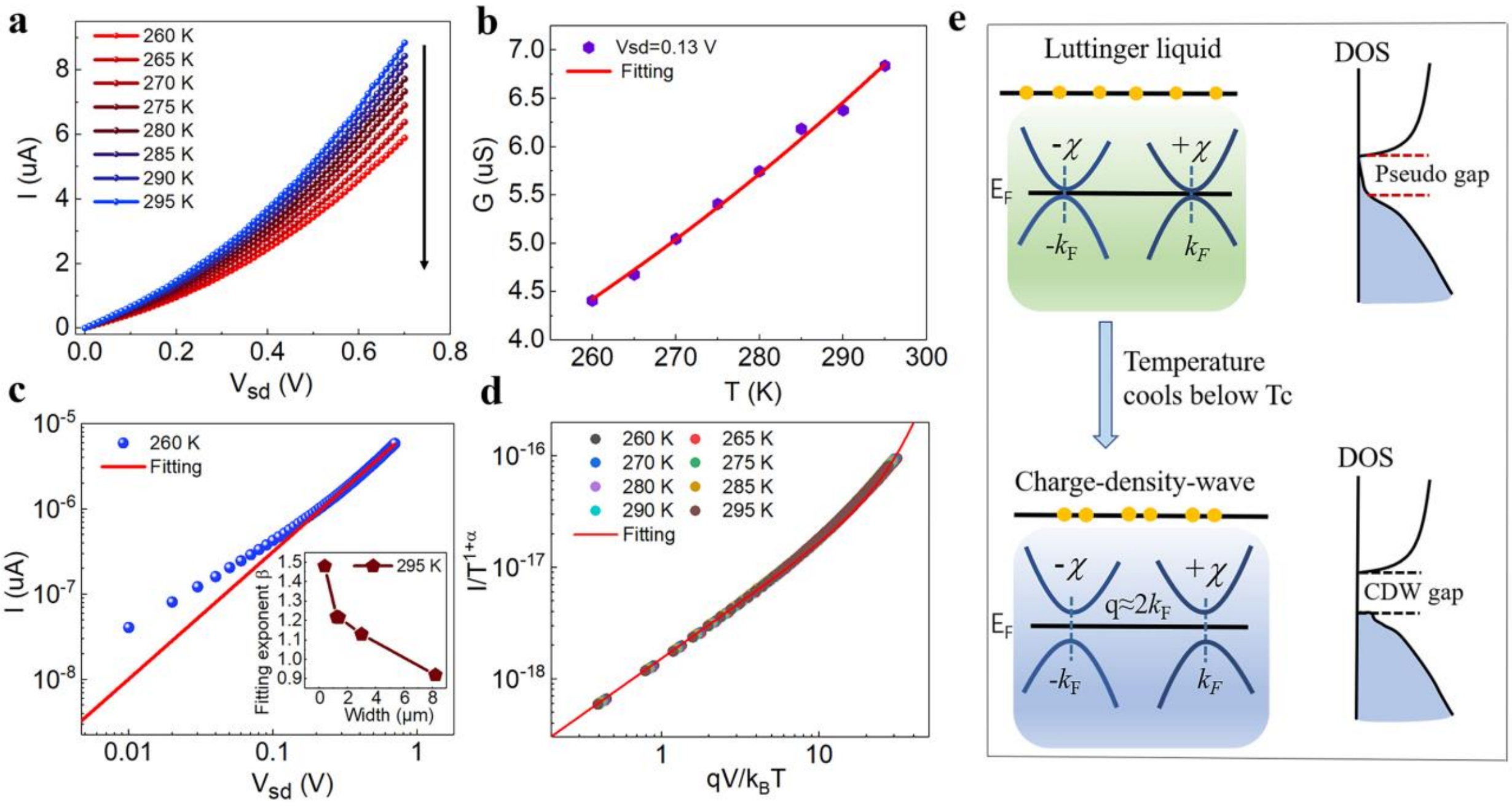

**Figure 4.** Luttinger liquid transport of $(TaSe_4)_2I$. a) Typical *I-V* curves measured under the temperature from 295 K to 260 K. b**)** Power law relation of conductance G with temperature ($G \sim T^{\alpha}$) at bias voltage of 0.13 V, obtaining a fitting exponent α = 3.47 from the red fitting curve. c) Power law relation of the *I-V* curve under 260 K, with a fitting exponent β =1.49 (Insert figure illustrates the sample width dependent fitting exponent β under 295 K). d) The output transport data and the corresponding curve plotted by $I/T_{1+\alpha}$ against $qV/k_BT$. e) Schematic band structure and density of state (DOS) corresponding to CDW and LL state.

## 2.5. Characterization of Joule Heating During the Photo Driven CDW Transition and Overall Photodetection Performance Evaluation

According to former literatures[10, 26], the photoresponse of quasi-1D CDW transition has two types of opposite behavior. One type is for photoresponse at temperatures far below $T_p$, the threshold electric field $E_T$ is enhanced under light illumination [10], which cannot be used for photodetection. The other type is the reduced $E_T$ under light illumination usually occurs at relatively high temperature[26], which is similar to our result. For the reduced $E_T$ condition, including in 2D CDW materials, the underlie mechanism is whether the intrinsic interplay of photogenerated electron holes with CDW electron lattice or the extrinsic Joule heating is still under debate up to date, while previous studies had indirectly measured or estimated the heating of the process[5b, 7a, 26b]. To clarify the mechanism clearly in our system, we did the direct in-situ time-resolved heating measurement on our device. Firstly, for bulk sample in dark *I-V* measurement, we could see when the voltage is biased just beyond $V_T$* (~ 0.1 s), the temperature fast rises by 90 K in 1.36 s (Figure S12a), and gradually rises in metallic state, finally goes down as the bias voltage decreases. We also estimated a temperature rise about 30 K on nanoplate sample used in in-situ Raman measurement, according to the temperature dependent Raman peak shift (Figure S12b). Secondly, for bulk sample photoresponse characterization, we test the process under 3 V bias voltage as an example. As the Figure 5a shows, when the laser illuminates, the temperature rises, but lags behind the ESCJ phenomenon for about 0.17 s, while before ESCJ region, the temperature only warms up ~ 1K, indicating that the main driving mechanism for ESCJ is not a laser heating or electric induced thermal effect. By such quantitative evaluation of the thermal effect in the photoelectric-field-induced CDW transition, it is demonstrated that Joule heating plays a secondary role in the CDW to metal transition, while it dominants the metal to CDW transition when the voltage sweeps back, which agrees with the previous report[5b]. The whole process is illustrated in Figure 5c. When a bias voltage applied to $(TaSe_4)_2I$, the “check board” like CDW pinning potential wells are inclined, equivalently the potential hight is slightly reduced, as is schematically shown in Figure 5c. I and 5c. II. Corresponding to the thermal measurement result, the light illumination does not trigger the sliding of CDW immediately, nevertheless, it either does not generate substantial heating to stimulate the transition to normal metal state, since the actual temperature increase is only ~1 K. The light illumination elevates the CDW electron energy further, or equivalently increases the phase slip probability of the CDW electron lattice, as shown in Figure 5c.III. Thereafter, in a period of time, the photoactivated electrons flow and cause local heat to stimulate the electron avalanche, and finally, the ESCJ occurs, as shown in Figure 5c.IV. Only after the ESCJ, substantial heating arises and the sample temperature has a much larger increasement.

Besides, commonly used HgCdTe devices involve complex manufacturing procedure, which is difficult to integrate with other materials on the silicon chip. Owing to the layered structure, the exfoliated thin flakes could be compatible with different substrates or hybridized with other material to realize more functionalities. Unlike BP or b-As which degrades quickly in air, our bulk $(TaSe_4)_2I$ devices also show great stability in ambient dry conditions over 6 months. For mid-infrared photodetection, our $(TaSe_4)_2I$ photodetectors simultaneously own both high responsivity, light on/off ratio and stability above liquid nitrogen temperature (Figure 5d). Although some MIR photodetectors based on photogating mechanism show relatively high responsivity (i.e. Te, InSb) worked at room temperature, however due to their small gain (compared to other mechansim in Figure 5d), they would possibly still suffer low light on/off ratio even if working at low temperature. Furthermore, the performance is limited by their bandgap, while the measured broadband transmittance spectrum demonstrates the optical absorption of (TaSe4)2I beyond 10 µm (FigureS2b), indicating that it also has the potential for realizing high photo responsivity and on/off ratio in long-infrared band. In additon, compared with currently reported mid-infrared photodetectors or APDs, it is also competitive when considering other parameters such detectivity and gain (Table S1 and S2). Though the response time is relatively slow (at millisecond range) at present stage for our devices, there is possibility to achieve much faster response time in optimized test condition, i.e. using under femtosecond laser radiation[7a].

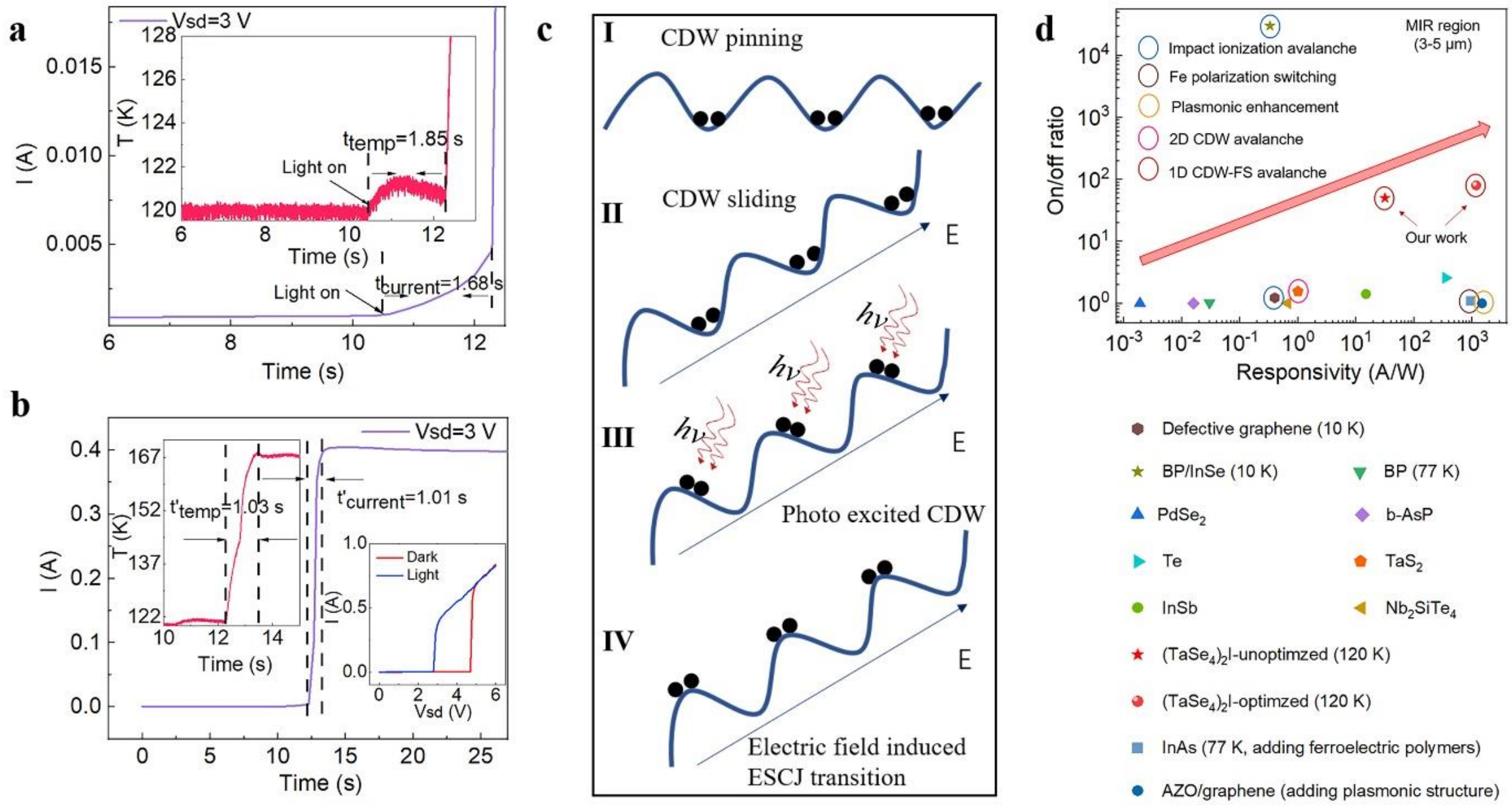

**Figure 5.** Characterization of Joule heating during the photo driven CDW transition and photodetection performance evaluation. a,b) In-situ time-resolved MIR photo-electric (~ light power density of 1.78 mW/mm$^2$) heating measurement of ESCJ phenemonon in short time sacle and long time scale. The right insert figure of b) is the *I-V* curves of the measured optimized high-quality sample (Sample2). c) The schematic mechanisim of photorespose driven by ESCJ in terms of electron energy activation (the electric field "E" in the figure is $E_T < E < E_T^*$). d) Photo responsivity and on/off ratio for reported low-dimensional materials based mid-infrared photodetectors[4, 7a, 27] . The devices work without any treatment unless it is specially labled. The working temperature is room temperature unless it is specially labled. Photodetectors based on photo-thermoelectric and photovoltaic mechanism are not included here for their usually low photo responsivity (below 1 A/W).

## 3. Conclusion

To summarize, we have investigated the CDW phase transition and the associated ESCJ phenomenon in quasi-1D $(TaSe_4)_2I$ at variable temperatures. The concurrent occurrence of a 1D structural phase transition is confirmed by in situ Raman spectroscopy measurement and DFT calculation. Compared to well-known CDW material quasi-2D $TaS_2$, $(TaSe_4)_2I$ photodetector are uncovered with several orders of magnitude higher MIR photo responsivity of 1.18 $\times 10^3$ A/W and larger light on-off ratio of 80 since the presence of "Fröhlich superconductivity" state and pseudogap respectively. Working above liquid nitrogen temperature, $(TaSe_4)_2I$ photodetector has outstanding photo responsivity, on/off ratio and stability within MIR range, demonstrating great potential in MIR detection. Our work discloses the important role of the dimensionality in quantum correlated phase transition, which also sheds light on the way of seeking their advanced photoelectronic application.

## 4. Experimental Section

**Materials Synthesis**: High-quality single crystals of $(TaSe_4)_2I$ were synthesized by chemical vapor transport (CVT) method in a sealed quartz tube. The high-purity Ta(4N), Se(4N) and I(4N) were mixed in chemical stoichiometry (defect-unoptimzed) or with an excess of I (defect-optimzed) sealed in an evacuated quartz tube which inserted into a furnace with a temperature gradient of 500 to 400 °C with the educts in the hot zone. After two weeks, needle-like crystals were grown with the typical size of ~1 mm × 150 μm. The thin (TaSe4)2I nanosheets were synthesized by liquid phase exfoliation (LPE) method for HRTEM and broadband aborsption characterization. Several bulk (TaSe4)2I crystal were added to10ml alcohol (30%) and subjected to bath sonication at 300 W for 2 hours. After centrifuging at 8000 rpm for 30 minutes to remove the deposit. Then thin $(TaSe_4)_2I$ nanosheets were obtained.

**Device fabrication:** For fabrication of the bulk crystal device, two 5 nm Ti and 100 nm Au-electrodes were deposited in sequence on the $(TaSe_4)_2I$ crystal flakes through thermal evaporation on freshly cleaved surface by using a mask to keep different distance between electric electrodes, and thin gold wires were soldered to the Au film areas by covering the silver paint completely to ensure the homogeneous injection of current. For fabrication of the thin crystal device, traditional Scotch tape exfoliation method were used. Electrode patterns were defined by standard electron beam lithography. Metal electrodes (10 nm Ti/200 nm Ag/100 nm Au) were deposited by thermal evaporator in PVD system (K.J.Lesker Nano 36).

**Electrical and photo response measurement:** The current–voltage (I–V) measurements were performed under current driving mode and voltage driving mode along the [110] direction. The samples were loaded in a closed cycle optical cryostat (SHI-4-1, Janis Ltd.) for room and low temperature measurements. The step dc electrical signals were provided and measured using a Keithley 2450 sourcemeter. Magnetic electric transport measurements were performed in in a closed cycle cryostat with a rotate sample chip carrier (Oxford Instruments Inc.). For wavelength-dependent photocurrent measurements, different continuous-wave solid-state lasers (Changchun New Industries Optoelectronics Technology Ltd. λ = 532, 635, 1064 nm) and two mid-IR continuous wave quantum cascade laser (λ =4.64 μm and λ =4.73 μm) were used as light sources. The incident light power illuminated on the device was monitored by calibrated power meters. For in-situ time-resolved photo-electric heating measurement, an E-type thermocouple was put near the channel of the sample to measure the local temperature.

**In-situ Raman and optical absorption measurements:** For room temperature Raman measurement, confocal Raman spectroscopy was performed (alpha300R, WITech Ltd.) on freshly cleaved $(TaSe_4)_2I$ crystals under a 100x objective in the room temperature using a grating of 1800 g/mm. In order to avoid laser-induced damage of the samples, the Raman spectrum were recorded at low power level (P ~ 500 μW). For low temperature Raman measurement, the sample was fixed on a sample probe with a 50x objective loaded in a closed cycle cryostat (Oxford Instruments Inc.). A narrow line-width laser of λ = 532 nm (Changchun New Industries Optoelectronics Technology Co. Ltd, MSL-FN-532) was used as laser source. A rotatable half waveplate was used to change the incident light polarization. The Raman spectrum was collected by liquid nitrogen refrigeration spectrometer (Princeton Instruments Inc.) using a grating of 1800 g/mm. For broadband optical absorption analysis, the transmittance spectra were measured by a UV-NIR spectrometer (Agilent Cray 5000) and a FITR (Vertex 70) spectrometer under at room temperature.

**DFT calculations:** The $(TaSe_4)_2I$ band structure calculations were performed using the Vienna ab initio simulation package (VASP)[28] with the projector-augmented wave (PAW)[29] method. The generalized gradient approximation (GGA) with revised Perdew, Burke, and Ernzerhof realization for solids (PBEsol)[30] as used for the exchange-correlation functional. The energy cutoff was set above 400 eV and the force and energy convergence criteria were set to 0.001 eV/Å and $10^{-8}$ eV respectively. A 7 × 7 × 7 Γ-centered k-point mesh was used for the Brillouin zone sampling. The Raman spectrum calculation is performed by using Raman-sc software packages [31].

**Supporting Information**

Supporting Information is available from the Wiley Online Library or from the author.

**Acknowledgements**

L.J Li acknowledges the funding from National Key R&D Program of China (2019YFA0308602), National Science Foundation of China (91950205 & 12174336) and the Zhejiang Provincial Natural Science Foundation of China (LR20A040002), Z. A Xu acknowledges the funding from National Key R&D Program of China (2019YFA0308602) and National Science Foundation of China (11774305), Y.H Lu acknowledges the funding of National Key R&D Program of China (2019YFE0112000), the Zhejiang Provincial Natural Science Foundation of China (LR21A040001) and the National Natural Science Foundation of China (11974307). J.L Li and H. Bai contributed equally to this work.

**Conflict of Interests**

The authors declare no conflict of interests.

**Reference**

[1] a)A. Rogalski, *Progress in Quantum Electronics* **2003**, 27, 59; b)E. Theocharous, J. Ishii, N. Fox, *Infrared Physics & Technology* **2005**, 46, 309; c)E. Michel, J. Xu, J. D. Kim, I. Ferguson, M. Razeghi, *IEEE Photon. Technol. Lett.* **2002**, 8, 673.
[2] A. Kerlain, G. Bonnouvrier, L. Rubaldo, G. Decaens, Y. Reibel, P. Abraham, J. Rothman, L. Mollard, E. De Borniol, *J. Electron. Mater.* **2012**, 41, 2943.
[3] J. Li, A. Dehzangi, G. Brown, M. Razeghi, *Sci. Rep.* **2021**, 11, 7104.
[4] A. Ga O, J. Lai, Y. Wang, Z. Zhu, J. Zeng, G. Yu, N. Wang, W. Chen, T. Cao, W. Hu, *Nat. Nanotechnol.* **2019**.
[5] a)X. Ma, T. Dai, S. Dang, S. Kang, X. Chen, W. Zhou, G. Wang, H. Li, P. Hu, Z. He, Y. Sun, D. Li, F. Yu, X. Zhou, H. Chen, X. Chen, S. Wu, S. Li, *ACS Appl Mater Inter* **2019**, 11, 10729; b)C. Zhu, Y. Chen, F. Liu, S. Zheng, X. Li, A. Chaturvedi, J. Zhou, Q. Fu, Y. He, Q. Zeng, H. J. Fan, H. Zhang, W. J. Liu, T. Yu, Z. Liu, *ACS Nano* **2018**, 12, 11203.
[6] a)J. Renteria, R. Samnakay, C. Jiang, T. R. Pope, P. Goli, Z. Yan, D. Wickramaratne, T. T. Salguero, A. G. Khitun, R. K. Lake, *J. Appl. Phys.* **2014**, 115, 2520; b)G. Liu, E. X. Zhang, C. D. Liang, M. A. Bloodgood, T. T. Salguero, D. M. Fleetwood, A. A. Balandin, *IEEE Electr Device L* **2017**, 38, 1724; c)M. Yoshida, R. Suzuki, Y. Zhang, M. Nakano, Y. Iwasa, *Sci Adv* **2015**, 1, e1500606.
[7] a)D. Wu, Y. Ma, Y. Niu, Q. Liu, T. Dong, S. Zhang, J. Niu, H. Zhou, J. Wei, Y. Wang, Z. Zhao, N. Wang, *Sci Adv* **2018**, 4, eaao3057; b)X. Wang, H. Liu, J. Wu, J. Lin, W. He, H. Wang, X. Shi, K. Suenaga, L. Xie, *Adv Mater* **2018**, 30, 1800074; c)C. Dang, M. Guan, S. Hussain, W. Wen, Y. Zhu, L. Jiao, S. Meng, L. Xie, *Nano Lett.* **2020**, 20, 6725; d)L. Wang, J. Wang, C. Liu, H. Xu, A. Li, D. Wei, Y. Liu, G. Chen, X. Chen, W. Lu, *Adv. Funct. Mater.* **2019**, 29.
[8] P. Monceau, *Advances in Physics* **2012**, 61, 325.
[9] W. Wen, Y. Zhu, C. Dang, W. Chen, L. Xie, *Nano Lett.* **2019**, 19, 1805.
[10] a)N. Ogawa, A. Shiraga, R. Kondo, S. Kagoshima, K. Miyano, *Phys. Rev. Lett.* **2001**, 87, 256401; b)S. V. Zaitsev-Zotov, V. E. Minakova, *JETP Lett.* **2004**, 79, 550.
[11] a)S. V. Zaitsev-Zotov, V. E. Minakova, *Phys. Rev. Lett.* **2006**, 97, 266404; b)A. Tomeljak, H. Schafer, D. Stadter, M. Beyer, K. Biljakovic, J. Demsar, *Phys. Rev. Lett.* **2009**, 102, 066404.
[12] a) Mitrovic S, Grioni M, Perfetti L. *J. Electron Spectrosc. Relat. Phenom* **2002**, 127, 77-84; b)L. Kang, X. Du, J. S. Zhou, X. Gu, Y. J. Chen, R. Z. Xu, Q. Q. Zhang, S. C. Sun, Z. X. Yin, Y. W. Li, D. Pei, J. Zhang, R. K. Gu, Z. G. Wang, Z. K. Liu, R. Xiong, J. Shi, Y. Zhang, Y. L. Chen, L. X. Yang, *Nat Commun* **2021**, 12, 6183; c)B. Dardel, D. Malterre, M. Grioni, P. Weibel, Y. Baer, F. Lévy, *Phys. Rev. Lett.* **1991**, 67, 3144; d)R. Claessen, G.-H. Gweon, F. Reinert, J. Allen, W. Ellis, Z. Shen, C. Olson, L. Schneemeyer, F. Lévy, *J. Electron Spectrosc. Relat. Phenom.* **1995**, 76, 121; e) Grioni M , Perfetti L , Berger H , et al. Evidence for strong correlations in a 1D Peierls system. *Physica* **2002**, 312, 559-561.
[13] S. Ghosh, F. Kargar, N. R. Sesing, Z. Barani, T. T. Salguero, D. Yan, S. Rumyantsev, A. A. Balandin, *Adv. Electron. Mater.* **2022**, 2200860.
[14] J. Cheng, C. An, L. Li, L. Chen, Y. Cui, Q. Yan, Y. Yin, W. Zhou, Y. Peng, W. Wang, D. Tang, *Appl. Phys. Lett* **2021**, 119, 201909.
[15] a)W. Shi, B. J. Wieder, H. L. Meyerheim, Y. Sun, Y. Zhang, Y. Li, L. Shen, Y. Qi, L. Yang, J. Jena, P. Werner, K. Koepernik, S. Parkin, Y. Chen, C. Felser, B. A. Bernevig, Z.

Wang, *Nat. Phys.* **2021**, 17, 381; b)J. Gooth, B. Bradlyn, S. Honnali, C. Schindler, N. Kumar, J. Noky, Y. Qi, C. Shekhar, Y. Sun, Z. Wang, B. A. Bernevig, C. Felser, *Nature* **2019**, 575, 315.
[16] T. Sekine, T. Seino, M. Izumi, E. Matsuura, *Solid State Commun.* **1985**, 53, 767.
[17] a)V. Favre-Nicolin, S. Bos, J. E. Lorenzo, J. L. Hodeau, J. F. Berar, P. Monceau, R. Currat, F. Levy, H. Berger, *Phys. Rev. Lett.* **2001**, 87, 015502; b)J. E. Lorenzo, R. Currat, P. Monceau, B. Hennion, H. Berger, F. Levy, *J Phys Condens Mat* **1998**, 10, 5039; c)Y. Zhang, L. F. Lin, A. Moreo, S. Dong, E. Dagotto, *Phys Rev B* **2020**, 101, 174106.
[18] a)G. Mihály, P. Beauchêne, *Solid State Commun.* **1987**, 63, 911; b)I. A. Cohn, S. G. Zybtsev, A. P. Orlov, S. V. Zaitsev-Zotov, *JETP Letters* **2020**, 112, 88.
[19] M. Maki, M. Kaiser, A. Zettl, G. Grüner, *Solid State Commun.* **1983**, 46, 497.
[20] a)S. Cova, M. Ghioni, A. Lacaita, C. Samori, F. Zappa, *Appl. Opt.* **1996**, 35, 1956; b)J. Zheng, X. Xue, C. Ji, Y. Yuan, K. Sun, D. Rosenmann, L. Wang, J. Wu, J. C. Campbell, S. Guha, *Nat Commun* **2022**, 13, 1517.
[21] a)M. Bockrath, D. H. Cobden, J. Lu, A. G. Rinzler, R. E. Smalley, L. Balents, P. L. Mceuen, *Nature* **1999**, 397, 598; b)J. D. Yuen, R. Menon, N. E. Coates, E. B. Namdas, S. Cho, S. T. Hannahs, D. Moses, A. J. Heeger, *Nat Mater* **2009**, 8, 572; c)G. Yang, Y. Shao, J. Niu, X. Ma, M. Liu, *Nat Commun* **2020**, 11.
[22] E. Slot, M. Holst, H. Van Der Zant, S. Zaitsev-Zotov, *Phys. Rev. Lett.* **2004**, 93, 176602.
[23] H. P. Geserich, G. Scheiber, M. Dorrler, F. Lévy, P. Monceau, *Physica B+C* **1986**, 143, 198.
[24] S. V. Borisenko, A. A. Kordyuk, A. N. Yaresko, V. B. Zabolotnyy, D. S. Inosov, R. Schuster, B. Büchner, R. Weber, R. Follath, L. Patthey, H. Berger, *Phys. Rev. Lett.* **2008**, 100, 196402.
[25] T. Tom, S. Bryan, *Rep. Prog. Phys* **1999**, 62, 61.
[26] a)S. Zybtsev, V. Y. Pokrovskii, V. Nasretdinova, S. Zaitsev-Zotov, *J. Commun. Technol. Electron.* **2018**, 63, 1053; b)M. Kato, M. Koyano, S. i. Katayama, *Solid State Commun.* **2004**, 129, 545.
[27] a)Q. Liang, Q. Wang, Q. Zhang, J. Wei, S. X. Lim, R. Zhu, J. Hu, W. Wei, C. Lee, C. Sow, W. Zhang, A. T. S. Wee, *Adv. Mater.* **2019**, 31, e1807609; b)X. Chen, X. Lu, B. Deng, O. Sinai, Y. Shao, C. Li, S. Yuan, V. Tran, K. Watanabe, T. Taniguchi, D. Naveh, L. Yang, F. Xia, *Nat Commun* **2017**, 8, 1672; c)S. Yuan, C. Shen, B. Deng, X. Chen, Q. Guo, Y. Ma, A. Abbas, B. Liu, R. Haiges, C. Ott, T. Nilges, K. Watanabe, T. Taniguchi, O. Sinai, D. Naveh, C. Zhou, F. Xia, *Nano Lett.* **2018**, 18, 3172; d)M. Zhao, W. Xia, Y. Wang, M. Luo, Z. Tian, Y. Guo, W. Hu, J. Xue, *ACS Nano* **2019**, 13, 10705; e)L. Tong, X. Huang, P. Wang, L. Ye, M. Peng, L. An, Q. Sun, Y. Zhang, G. Yang, Z. Li, F. Zhong, F. Wang, Y. Wang, M. Motlag, W. Wu, G. J. Cheng, W. Hu, *Nat Commun* **2020**, 11, 2308; f)B. Y. Zhang, T. Liu, B. Meng, X. Li, G. Liang, X. Hu, Q. J. Wang, *Nat Commun* **2013**, 4, 1811; g)X. Zhang, H. Huang, X. Yao, Z. Li, C. Zhou, X. Zhang, P. Chen, L. Fu, X. Zhou, J. Wang, W. Hu, W. Lu, J. Zou, H. H. Tan, C. Jagadish, *ACS Nano* **2019**, 13, 3492; h)C.-H. Kuo, J.-M. Wu, S.-J. Lin, W.-C. Chang, *Nanoscale Res. Lett.* **2013**, 8, 327; i)S. Zhang, H. Jiao, X. Wang, Y. Chen, H. Wang, L. Zhu, W. Jiang, J. Liu, L. Sun, T. Lin, H. Shen, W. Hu, X. Meng, D. Pan, J. Wang, J. Zhao, J. Chu, *Adv. Funct. Mater.* **2020**, 30, 2006156; j)Y.-F. Yu, Y. Zhang, F. Zhong, L. Bai, H. Liu, J.-P. Lu, Z.-H. Ni, *Chin. Phys. Lett.* **2022**, 39, 058501.
[28] G. K. A, J. F. b, *Comp Mater Sci* **1996**, 6, 15.
[29] G. Kresse, D. Joubert, *Phys Rev B* **1999**, 59, 1758.
[30] J. P. Perdew, A. Ruzsinszky, G. I. Csonka, O. A. Vydrov, K. Burke, *Phys. Rev. Lett.* **2008**, 100, 136406.
[31] A. Fonari, S. Stauffer. vasp_raman.py. **2013**, https://github.com/raman-sc/VASP.

**Title: The Influence of Dimensionality on the Charge Density Wave Transition and Its Application on Mid-infrared Photodetection**

**Authors:** Jialin Li, Hua Bai, Yupeng Li, Junjian Mi, Qiang Chen, Wei Tang, Huanfeng Zhu, Xinyi Fan, Yunhao Lu*, Zhuan Xu*, Linjun Li*

**ToC figure:**

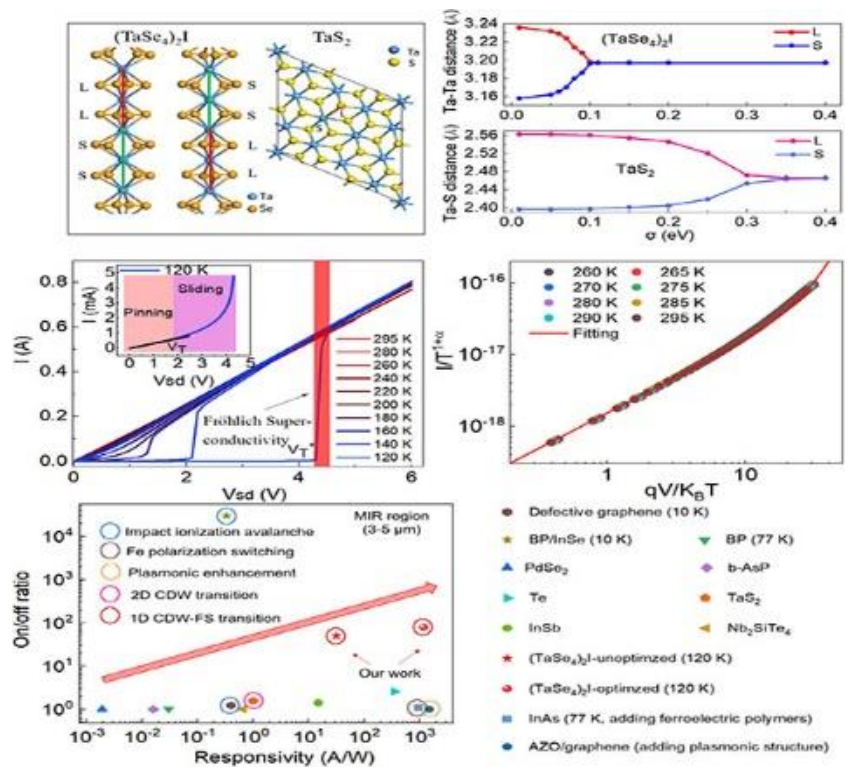

**Synopsis:**

Sharp 1D CDW transition in $(TaSe_4)_2I$ is experimentally and theoretically investigated, and applied for sensitive mid-infrared photodetection, the achieved photo responsivity and on/off ratio are superior than 2D CDW $TaS_2$ and most of other low-dimensional materials. Our work not only reveal the important role of the dimensionality in CDW transition, but also sheds light on the way for advanced mid-infrared application.

# Supporting Information

## The Influence of Dimensionality on the Charge Density Wave Transition and Its Application on Mid-infrared Photodetection

*Jialin Li, Hua Bai, Yupeng Li, Junjian Mi, Qiang Chen, Wei Tang, Huanfeng Zhu, Xinyi Fan, Yunhao Lu*, Zhuan Xu*, Linjun Li**

### 1. Material Characterization

The high-resolution transmission electron microscope (HRTEM) and selected area electron diffraction (SAED) image of the as-grown $(TaSe_4)_2I$ is shown in Figure S1a and Figure S1b respectively. The scanning electron microscope (SEM) image and energy dispersive spectrum (EDS) analysis result of the as-grown $(TaSe_4)_2I$ is shown in Figure S1c and its insert figure, the ratio of Ta:Se:I is 2.1:7.8:1.The corresponding element EDS mapping is shown in Figure S1d.

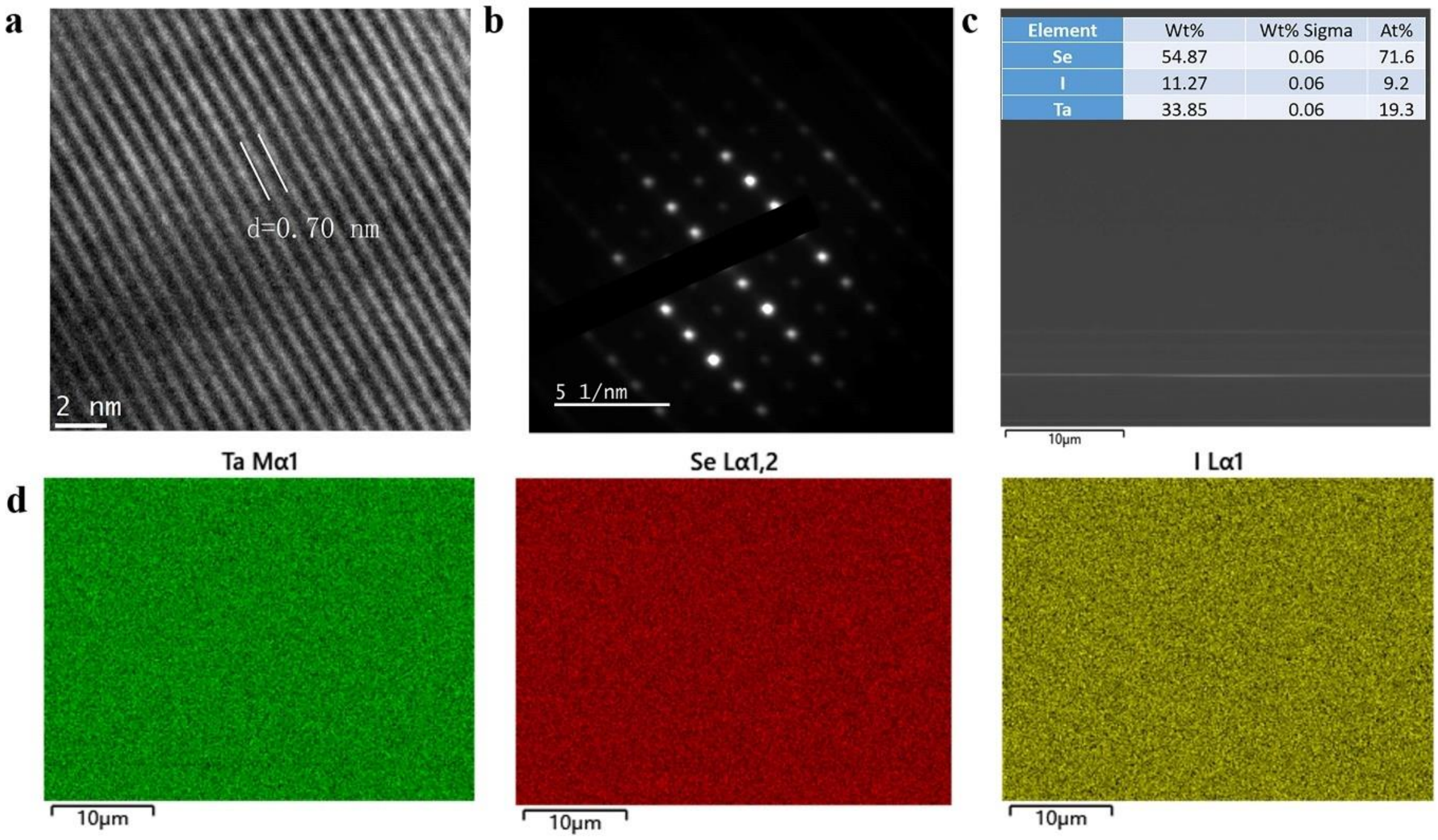

**Figure S1.** Material characterization of the as-grown $(TaSe_4)_2I$. (a) HRTEM image of $(TaSe_4)_2I$. (b) SAED image of $(TaSe_4)_2I$. (c) SEM image of $(TaSe_4)_2I$ surface. (d) EDS mapping of $(TaSe_4)_2I$.

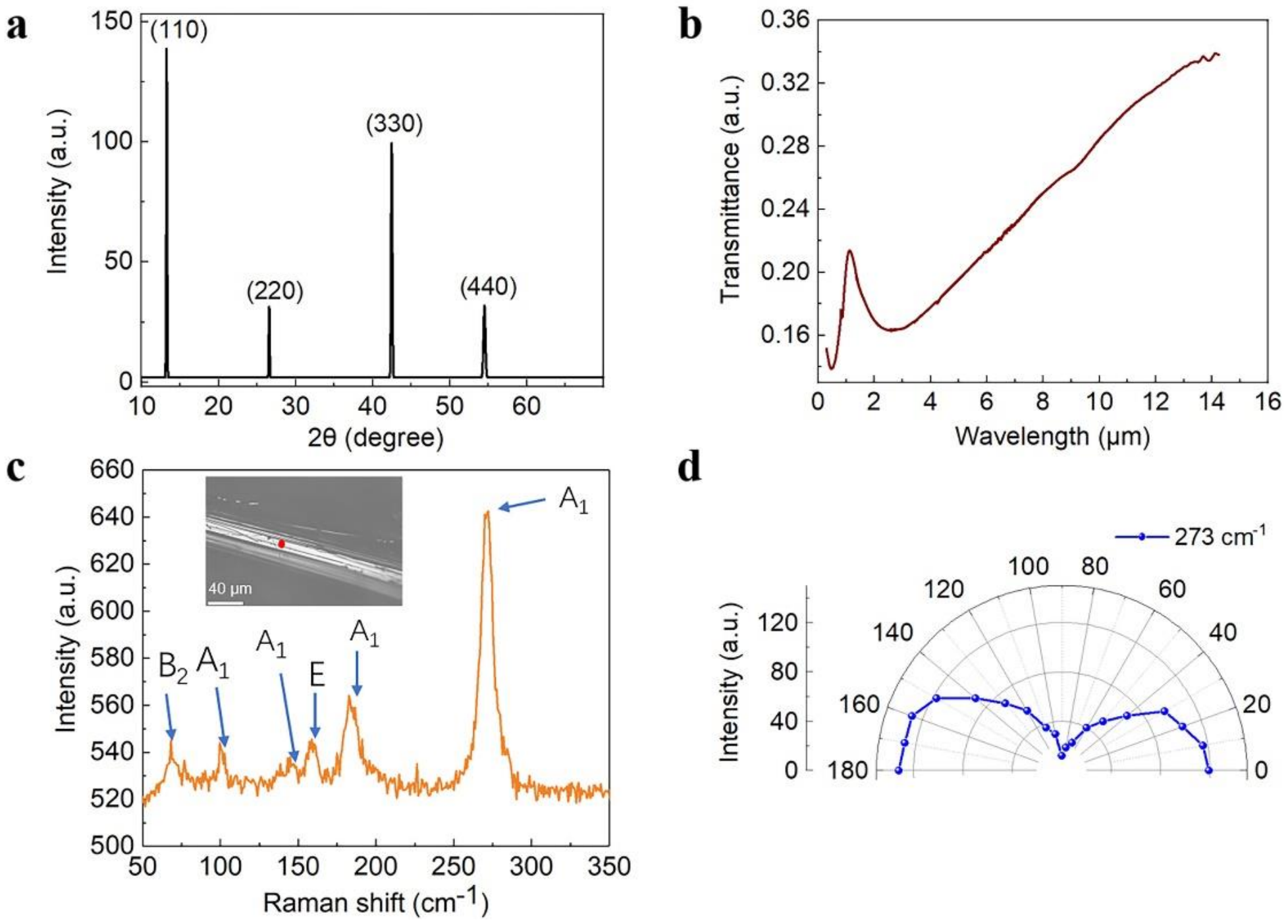

**Figure S2.** X-ray diffraction (XRD) spectrum, broadband transmittance spectrum, Raman spectrum and anisotropic property of (TaSe4)2I single crystal. (a) XRD spectrum. (b) Broadband transmittance spectrum from ultraviolet to long-infrared region. (c) Raman spectrum on a bulk crystal. Insert: optical image of the bulk (TaSe4)2I single crystal. (d) Polar plots of the Raman-peak intensity of 273 cm-1 as a function of polarization angle obtained by performing ploarized Raman spectrum.

**2. Nonlinear and Hysteresis Behavior at Different Temperature**

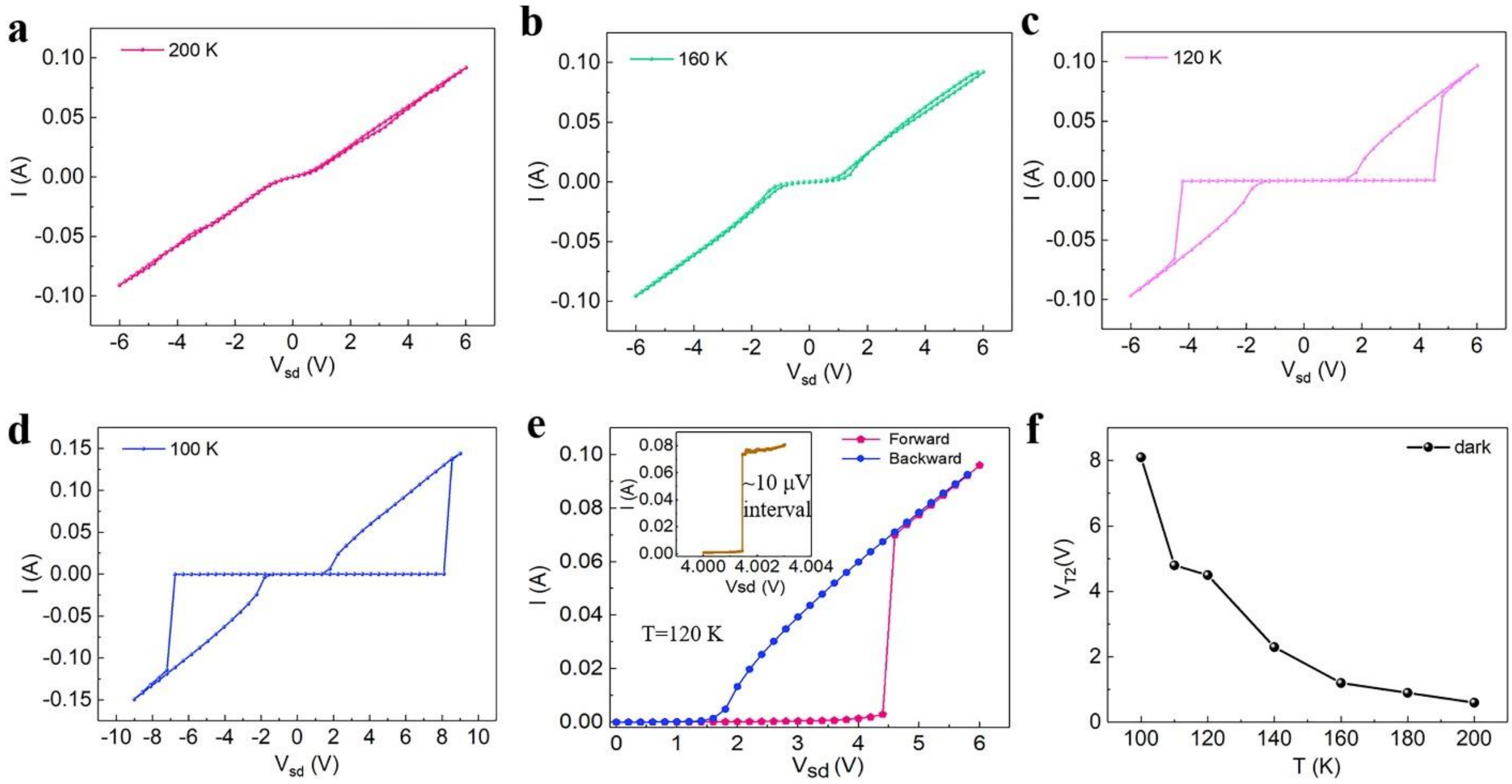

**Figure S3.** Nonlinear and hysteresis behavior at different temperature of device with the lateral size of 1 mm × 150 μm (sample 1). (a), (b), (c), (d), (e) *I*-*V* curves under different temperature. (f) Temperature dependent ESCJ threshold $V_{T*}$ (or called $V_{T2}$) variation.

## 3. Raman Calculation of Non-CDW Mode and CDW Mode of $(TaSe_4)_2I$

The Raman calculation is performed by using Raman-sc software packages. The σ is 0.01 for CDW Raman calculation and 0.2 for non-CDW Raman calculation respectively.

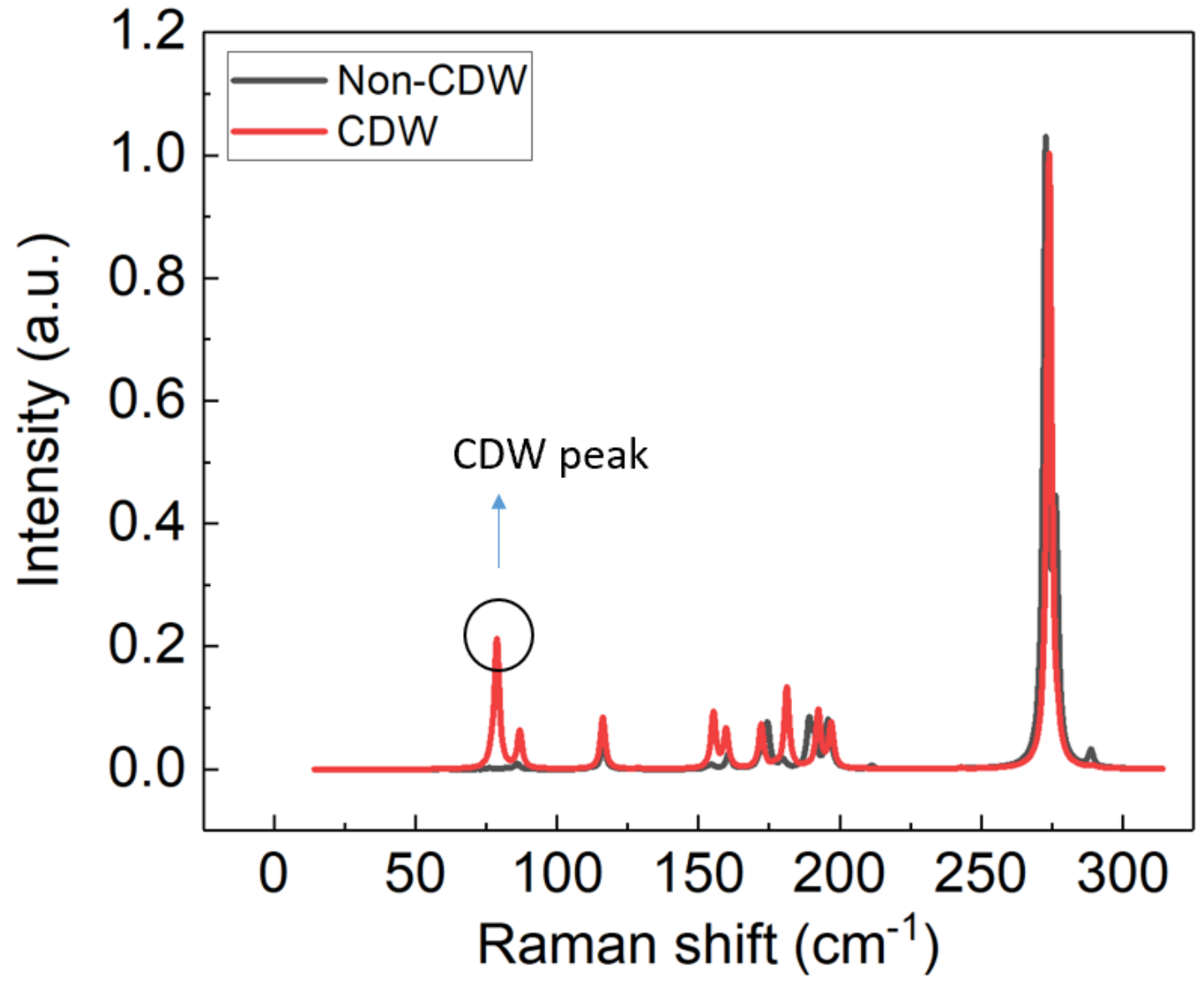

**Figure S4.** The Raman calculation of non-CDW mode and CDW mode of $(TaSe_4)_2I$.

## 4. The Density Functional Theory (DFT) Calculation

Figure S5 shows the band structure of $(TaSe_4)_2I$ with different value of parameter sigma (σ). All optimization results maintain the space group of I422. But the structure is very sensitive to changes of σ. Electronic band structure of $(TaSe_4)_2I$ with spin-orbit coupling (σ=0.01-0.05) shows semiconductor characteristics with a bandgap, while the bandgap decreases as the value of sigma increases. The band structure shows metal characteristics with near zero bandgap when the value of σ is above 0.1.

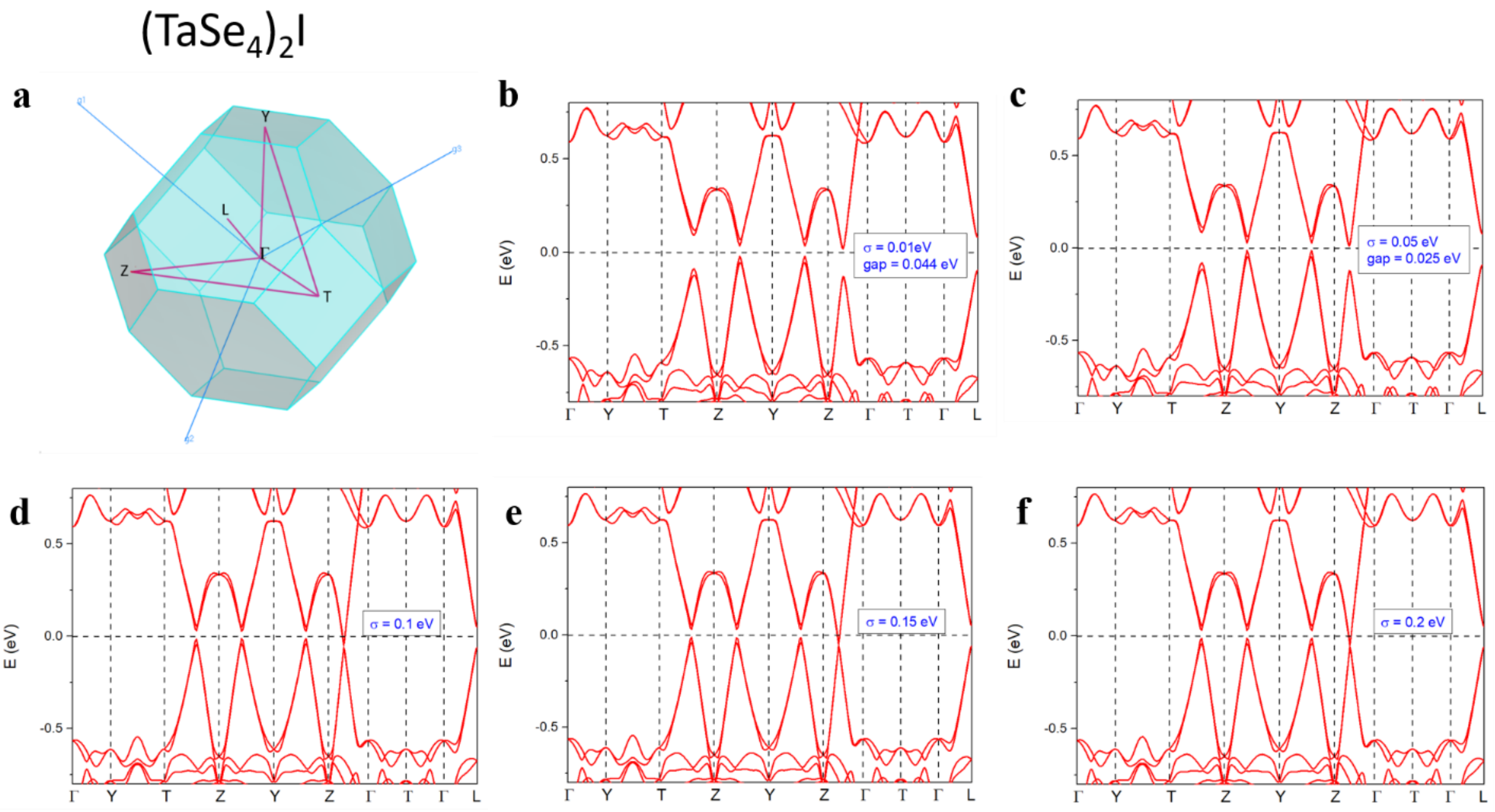

**Figure S5.** The band structure of $(TaSe_4)_2I$ with different value of parameter σ. (a) The bulk Brillouin zones of crystal structure. (b)-(c) Electronic band structure of $(TaSe_4)_2I$ with spin-orbit coupling (σ=0.01-0.05). (d)-(f) Electronic band structure of $(TaSe_4)_2I$ with spin-orbit coupling (σ=0.1-0.2).

## 5. Photo Reshaped CDW Transition at Different Temperature

Figure S6 shows the photo reshaped CDW sliding phenomenon at different temperature under a laser illumination of λ=532 nm. The spot size is about 1.6 mm, and the laser power is monitored by calibrated power meters, being fixed at 55 mW. As the temperature decreases, photo reshaped CDW transition phenomenon is more obvious.

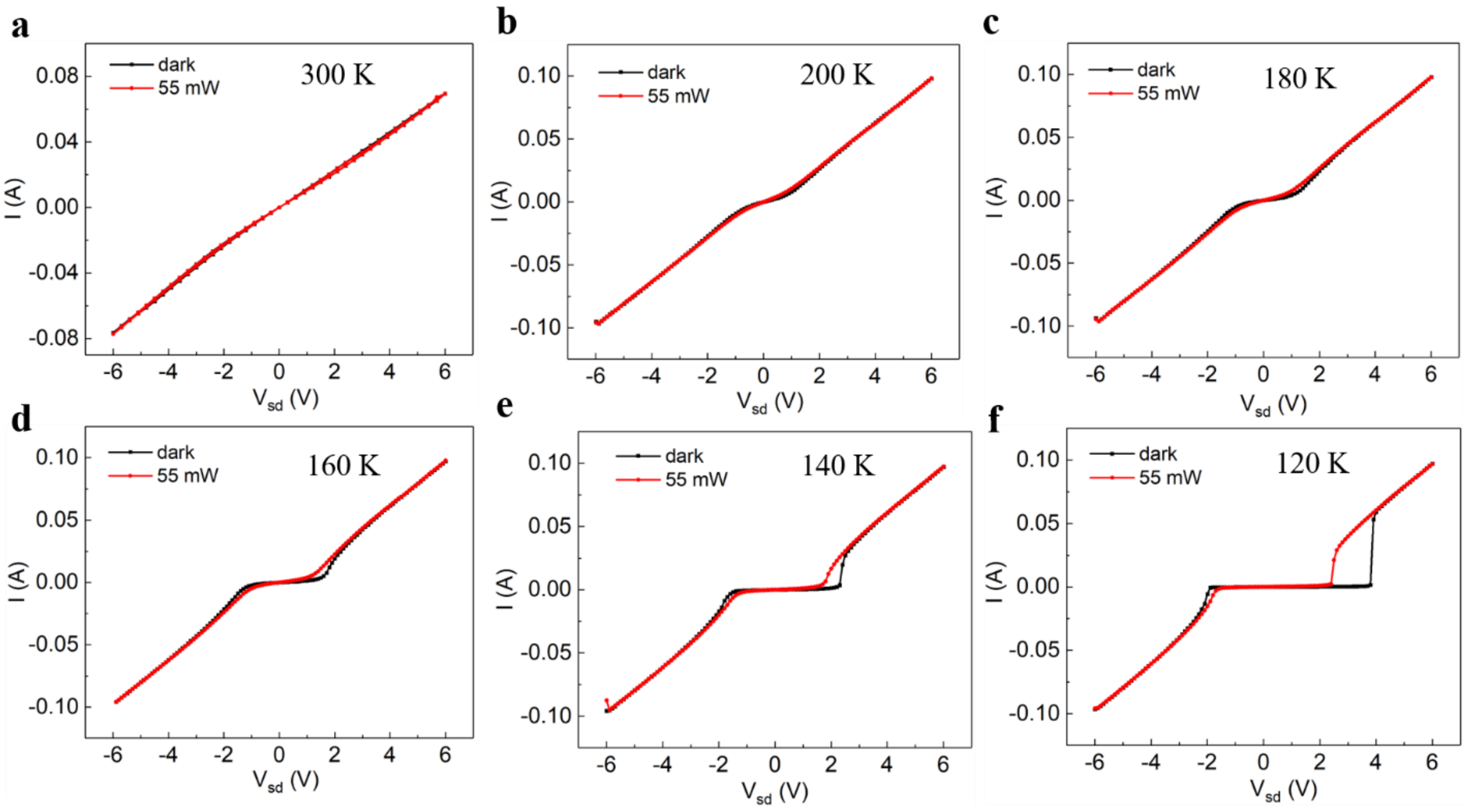

**Figure S6.** Photo reshaped CDW transition of sample1 at different temperature under a laser illumination of λ=532 nm. The laser spot size is about 1.6 mm. (a)-(f) is measured at 300 K, 200 K, 180 K, 160 K, 140 K and 120 K respectively.

## 6. Photo Response at Relatively Higher Temperature

Photo response of the two-probe device of sample 1 at 140 K is shown in Figure S7. As the photo power increase, the threshold voltage is decreased as expected at both positive and negative sweep, suggesting that the illumination can promote CDW depinning. The photo response speed can be further increased as the voltage rises near the threshold voltage with response time below 1 s. It is quite similar with $TaS_2$, which also realized high photo responsivity by light manipulating CDW state. Our result shows a higher photo responsivity of ~16 A/W than $TaS_2$ of near 4 A/W @ 532 nm.

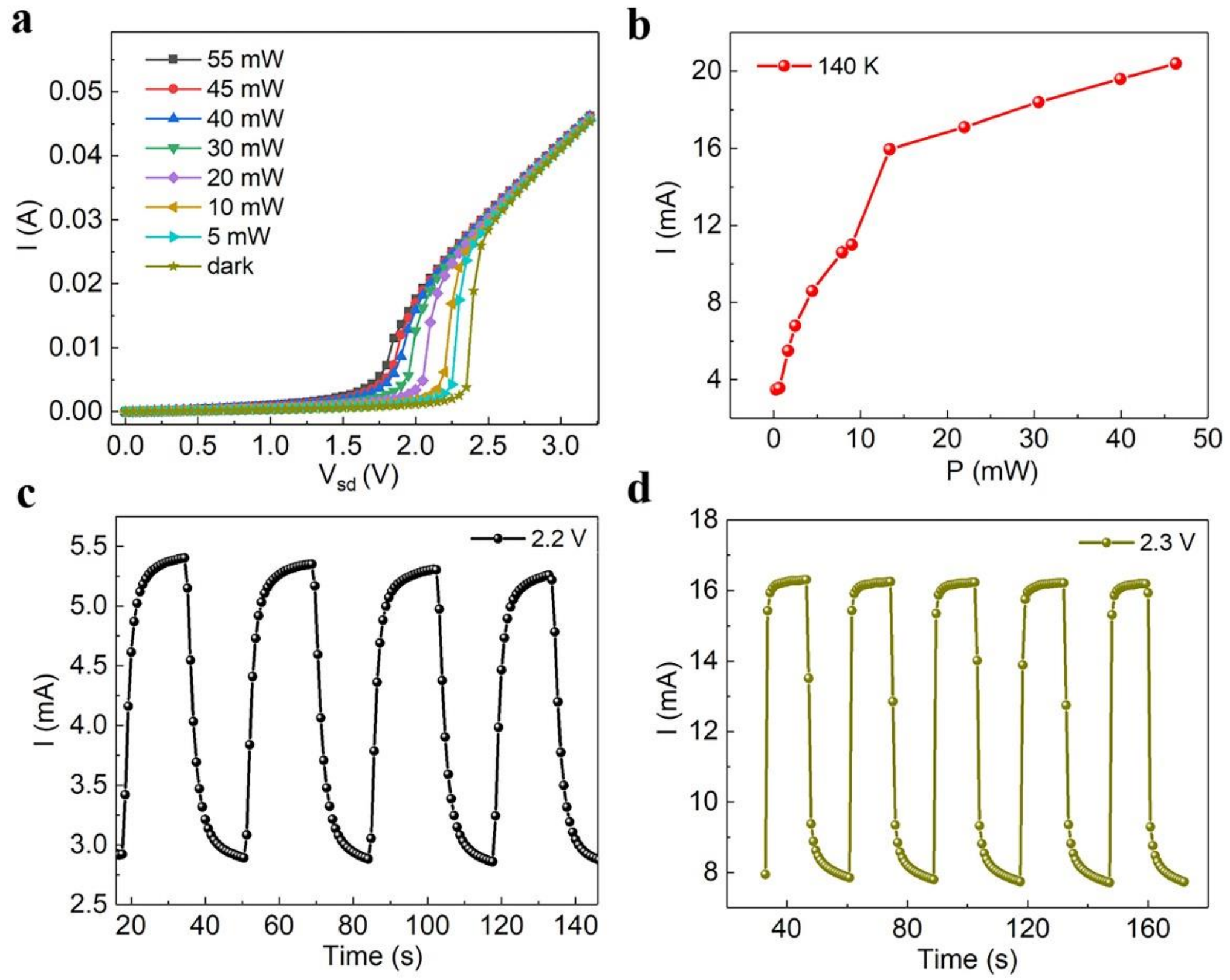

**Figure S7.** Photo response of sample 1 at 140 K. (a) photo reshaped CDW transition under positive bias voltage sweep. (b) Power dependent current response. (c) and (d) Photoresponse under bias votage of 2.2 V and 2.3 V respectively.

## 7. Power-dependent Photo Response Under a Laser Illumination of λ=532 nm

Figure S8a depicts the photo response under different laser illumination intensity of 532 nm with a fixed voltage of 2.5 V at 120 K. Here it can be divided into three processes. In region I, when the light intensity is below $P_c$, the CDW is not destroyed and photoexcited electrons contribute to the photocurrent which is proportional to light power. As the light intensity approaches $P_c$ in region II, CDW-Fröhlich superconductivity happens, Γ - Z gap suddenly closes, thus CDW state is turned into metal state, manifested by the huge elevation of the current. In region III, the photocurrent is contributed by the photo-thermoelectric effect.

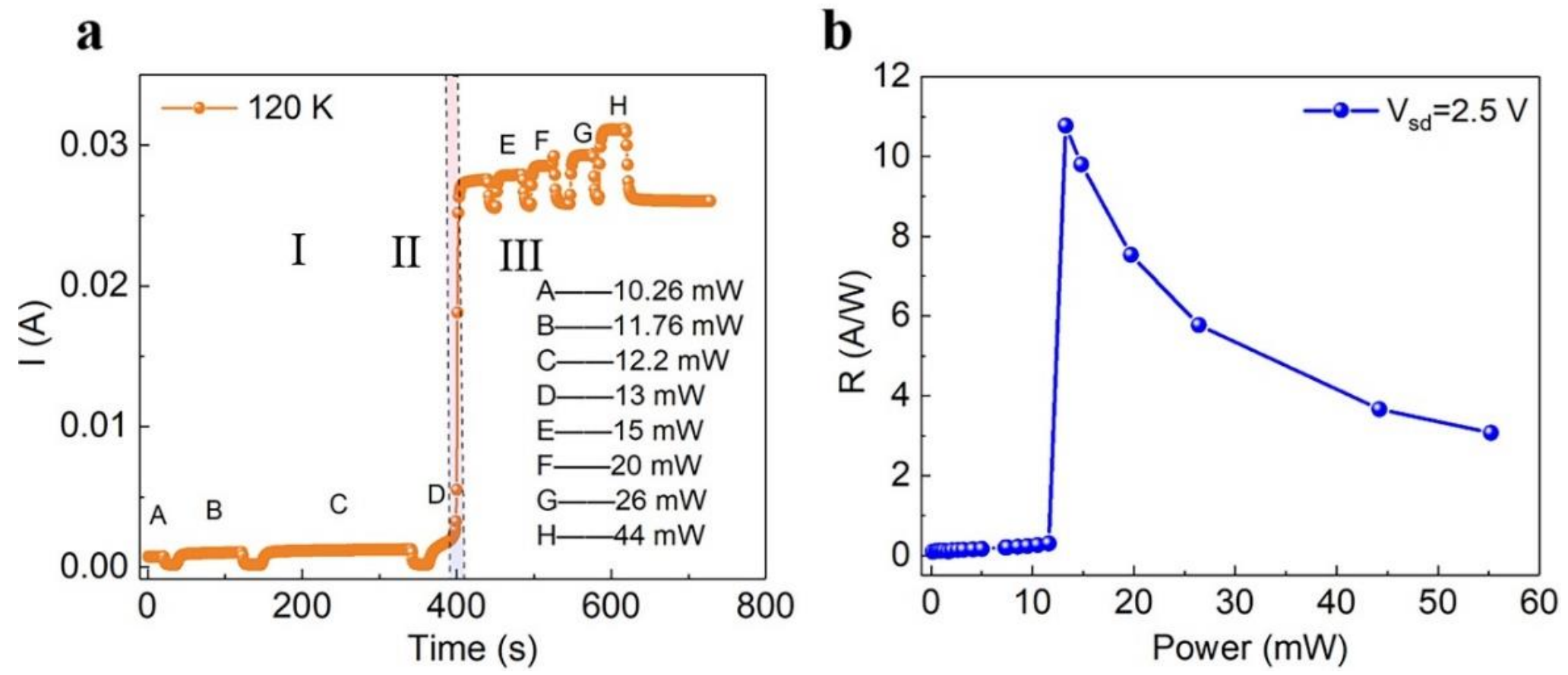

**Figure S8.** Photo response measurement and power dependent photo responsivity under a laser illumination of λ=532 nm at 120 K of sample 1. The spot size is about 1 mm here. (a) Photo response under different laser illumination intensity of λ=532 nm excitation with a fixed voltage of 2.5 V at 120 K. (b) The photoresponsivity under different light power under λ=532 nm excitation with bias voltage of 2.5 V.

**8. Power-Dependent Photo Response Under a Laser Illumination of λ=1064 nm**

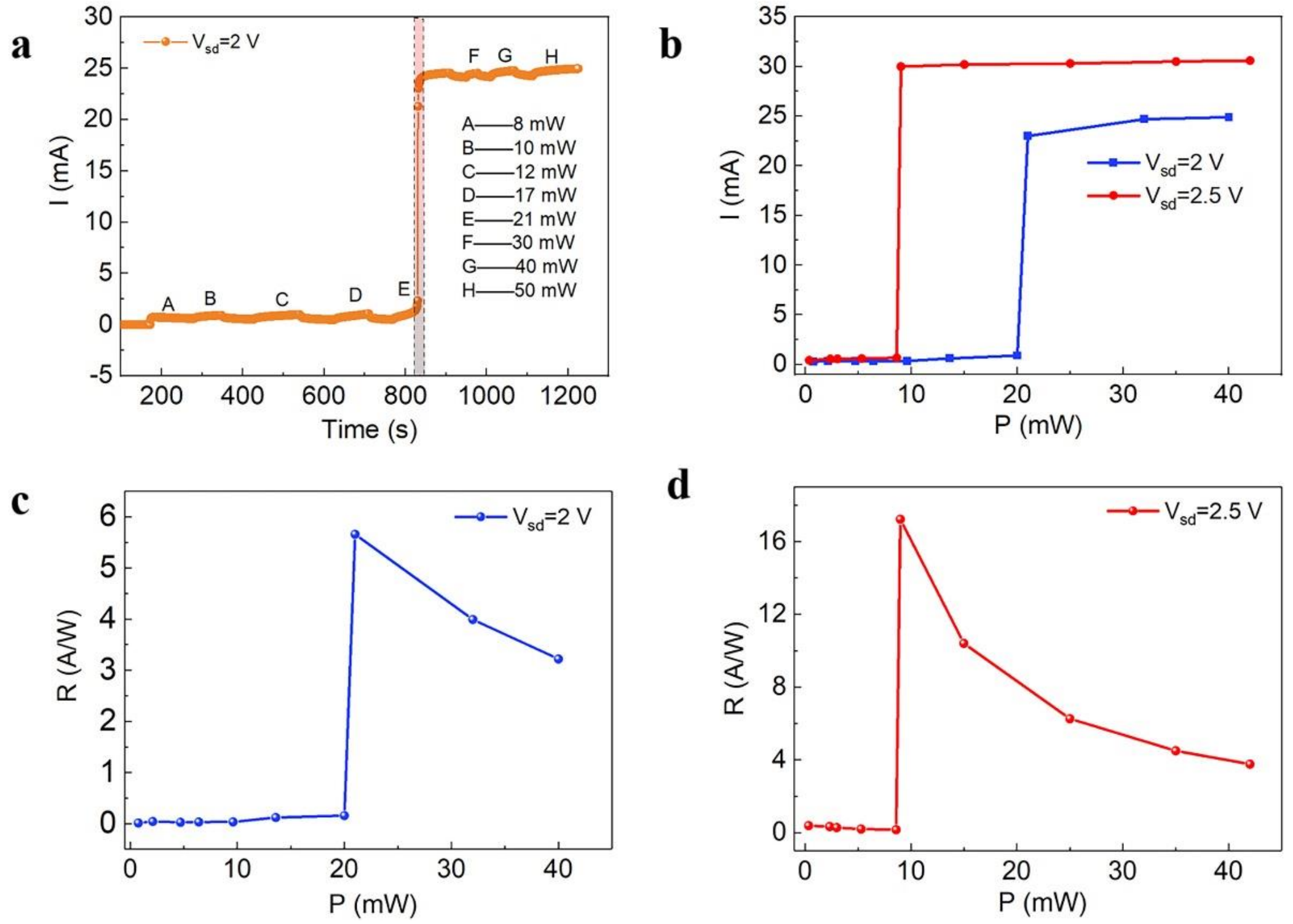

**Figure S9.** Photo-reshaped CDW transition phenomenon under laser illumination of λ=1064 nm at 120 K. The spot size is about 1 mm here.

**9. MIR Photo Response in a Thin $(TaSe_4)_2I$ Nanoribbon Device**

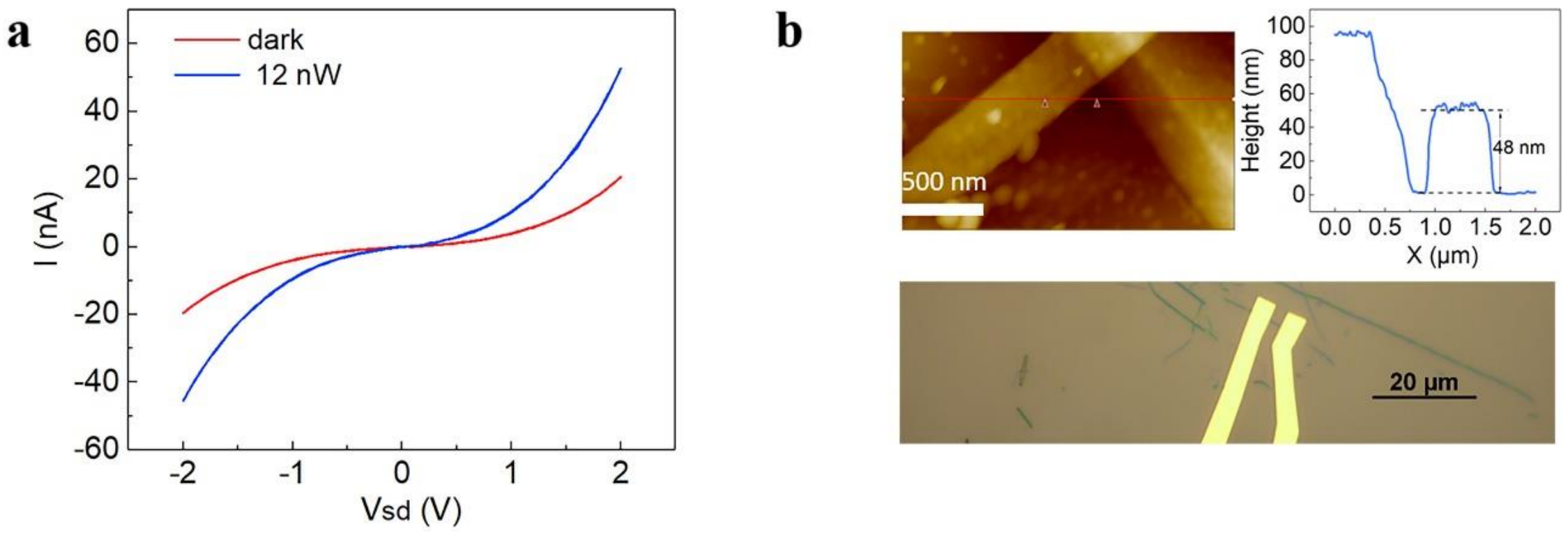

**Figure S10.** Photo response under a laser illumination of λ=4.64 μm at 120 K with the crystal lateral size of 2.4 μm ×0.4 μm ×48 nm. (a), *I-V* curves under low-electric field region. (b) Optical image, AFM image and height profile of the sample.

## 10. MIR Photo Response in a Thin $(TaSe_4)_2I$ Nanoplate Device

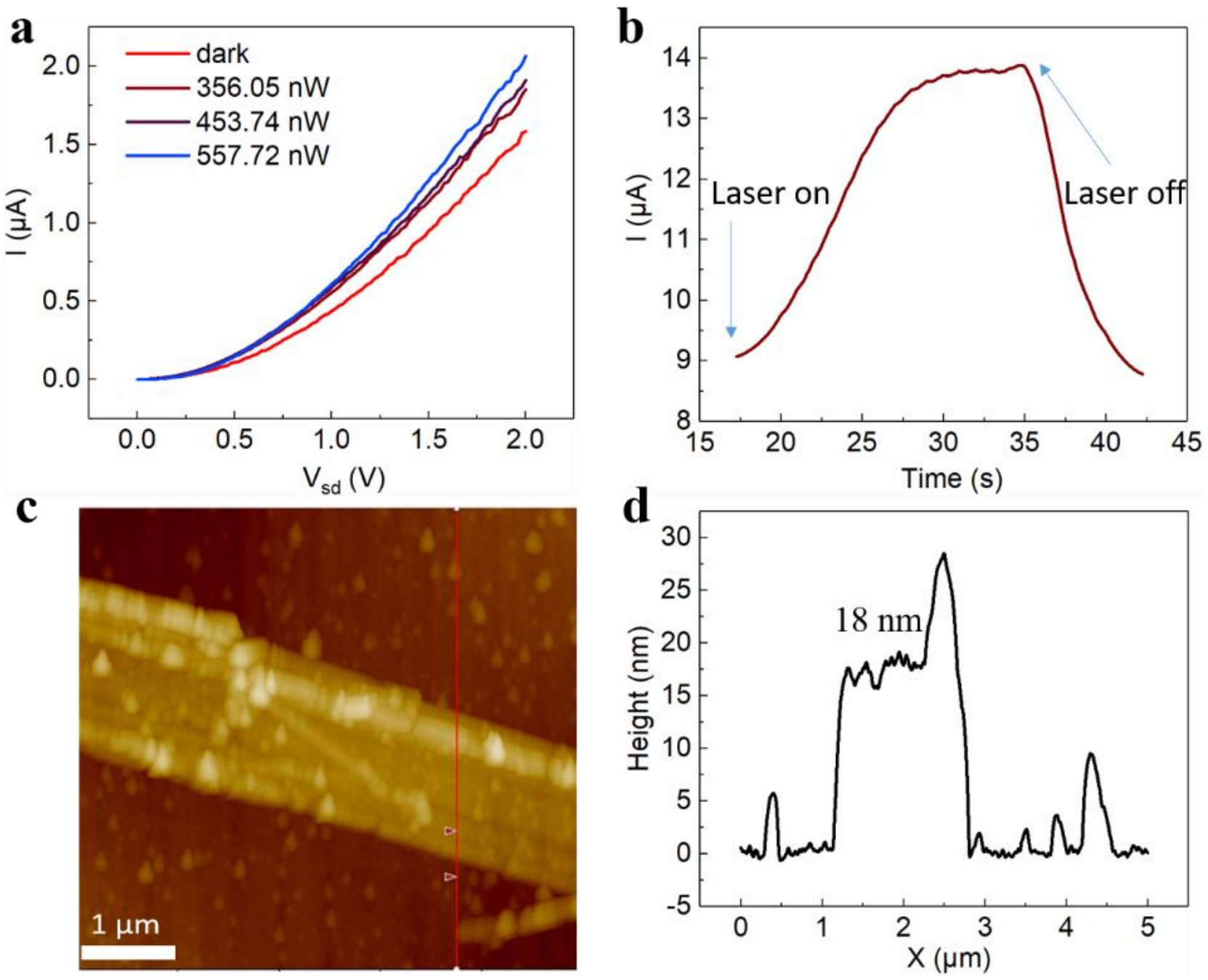

**Figure S11.** Photo response under a laser illumination of λ=4.64 μm at 120 K with the crystal lateral size of 7.5 μm × 1.3 μm ×18 nm (refer to S3). (a) I-V curves under low-electric field region with different illumination power. (b) Photo response under the higher bias voltage of 7 V with illumination power of 356.05 nW on the sample. (c) and (d) AFM image and height profile of the sample.

**11. In-situ Heat Measurement in Dark State of $(TaSe_4)_2I$.**

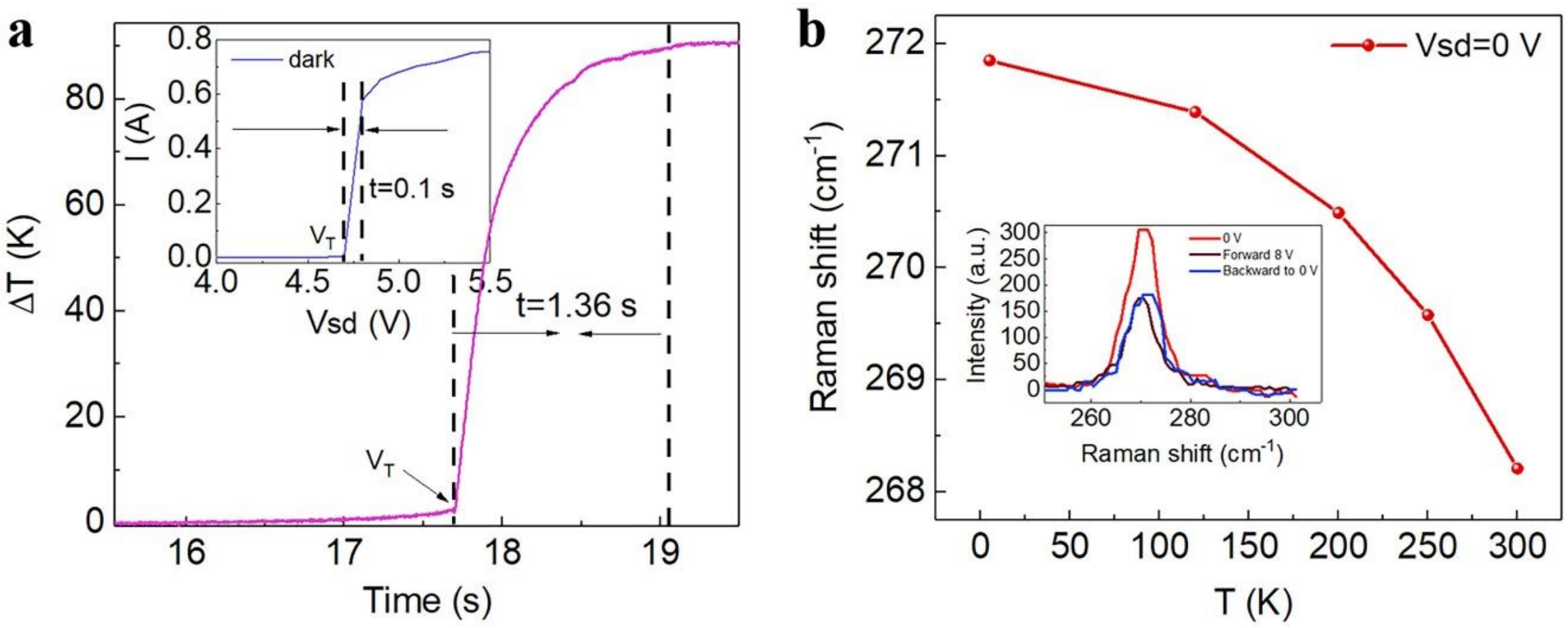

**Figure S12.** (a) Time-resolved heating measurement of dark state *I*-*V* on our high-quality sample (Sample 2). (b) Temperature dependent Raman peak shift on nanoplate sample used in in-situ Raman measurement.

**12. Luttinger Liquid Transport of $(TaSe_4)_2I$ Nano-thick Device with Width of 1.3 μm**

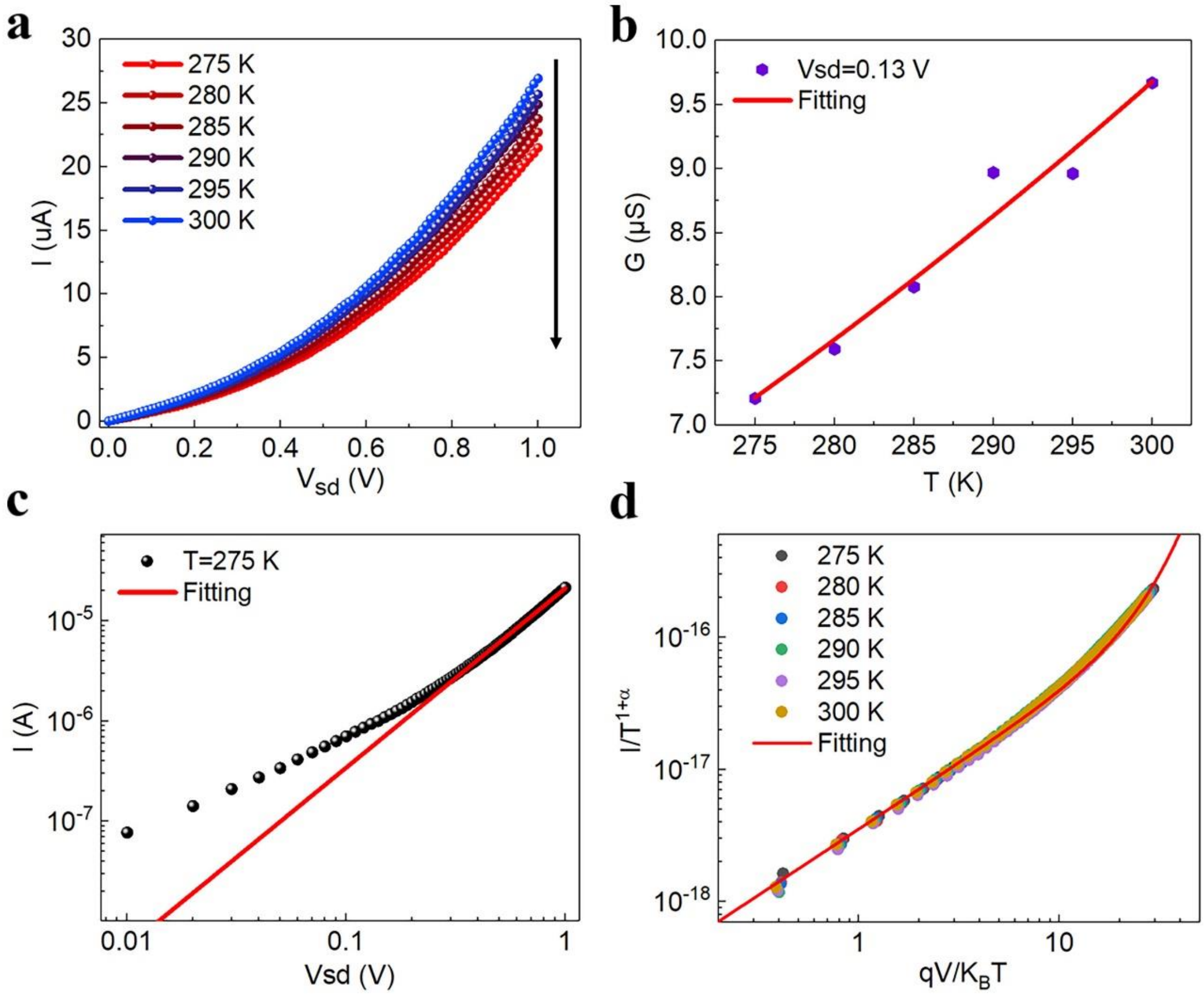

**Figure S13.** Luttinger liquid transport of $(TaSe_4)_2I$ nano-thick device with width of 1.3 μm.

**13. Luttinger Liquid Transport of $(TaSe_4)_2I$ Nano-thick Device with Width of 3 μm**

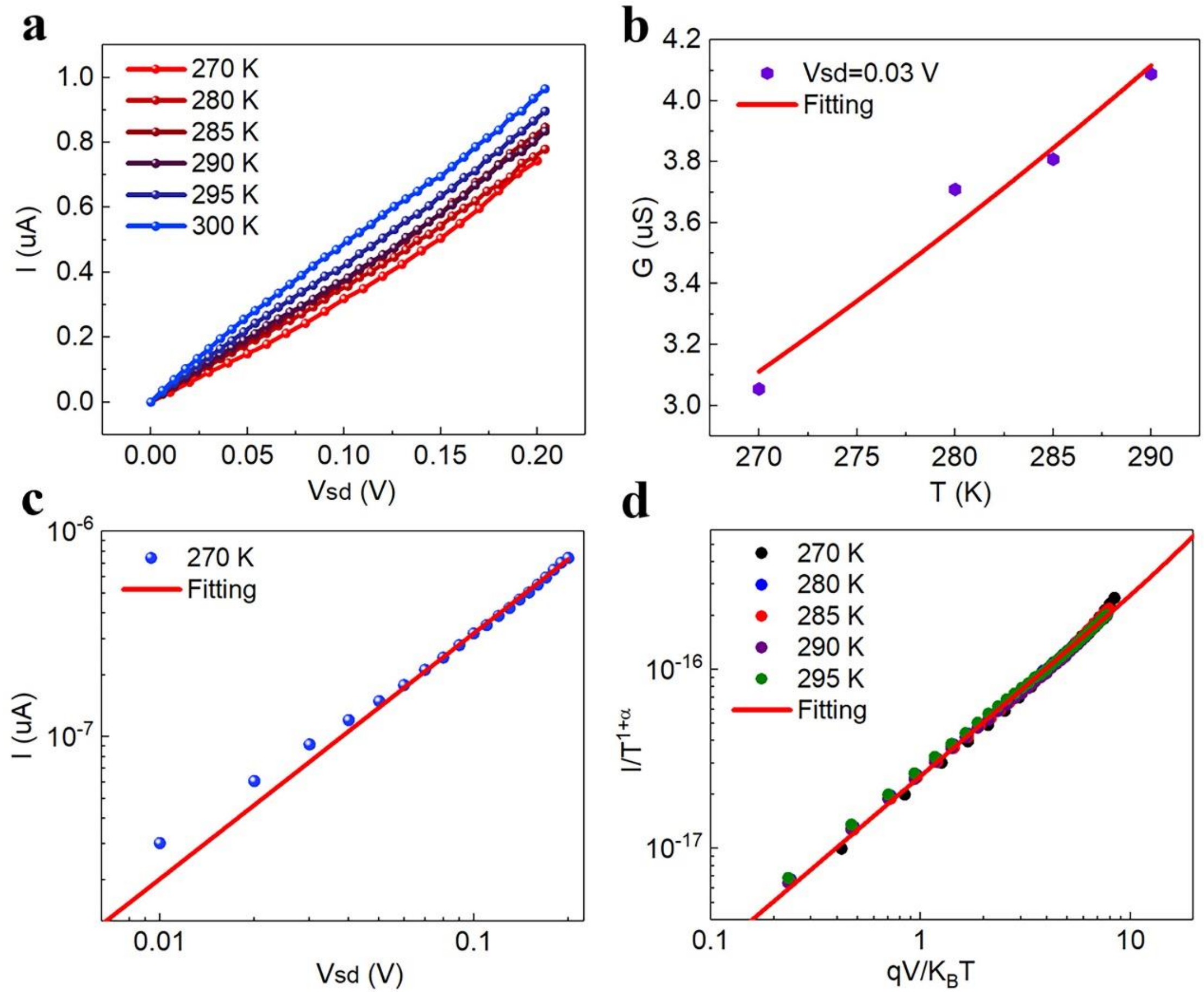

**Figure S14.** Luttinger liquid transport of $(TaSe_4)_2I$ nano-thick device with width of 3 μm.

**14. Photo Reshaped CDW Transition in 2D 1T-$TaS_2$ at Different Temperature**

At low temperature, $TaS_2$ also has metastable states, even if there is no metastable state at room temperature, as shown in Figure S15a. Also, due to the the small carrier concentration variation across the CDW transition, the dark current is still high even at low temperature, thus caused a poor photo response. While there is no metastable state and the CDW domain distribution is along the chain direction (the crystal is highly anisotropy) in $(TaSe_4)_2I$, it is very stable. In addition, due to the sharp carrier density variation across the CDW transition, the conductance can be changed several orders of magnitudes higher than $TaS_2$ and hence obtain much higher photo responsivity and light on/off ratio. In addition, due to the small *I-V* hysteresis of $TaS_2$, the photodetector can only be worked in a limited voltage range (＜0.5 V) as shown in Figure S15d.

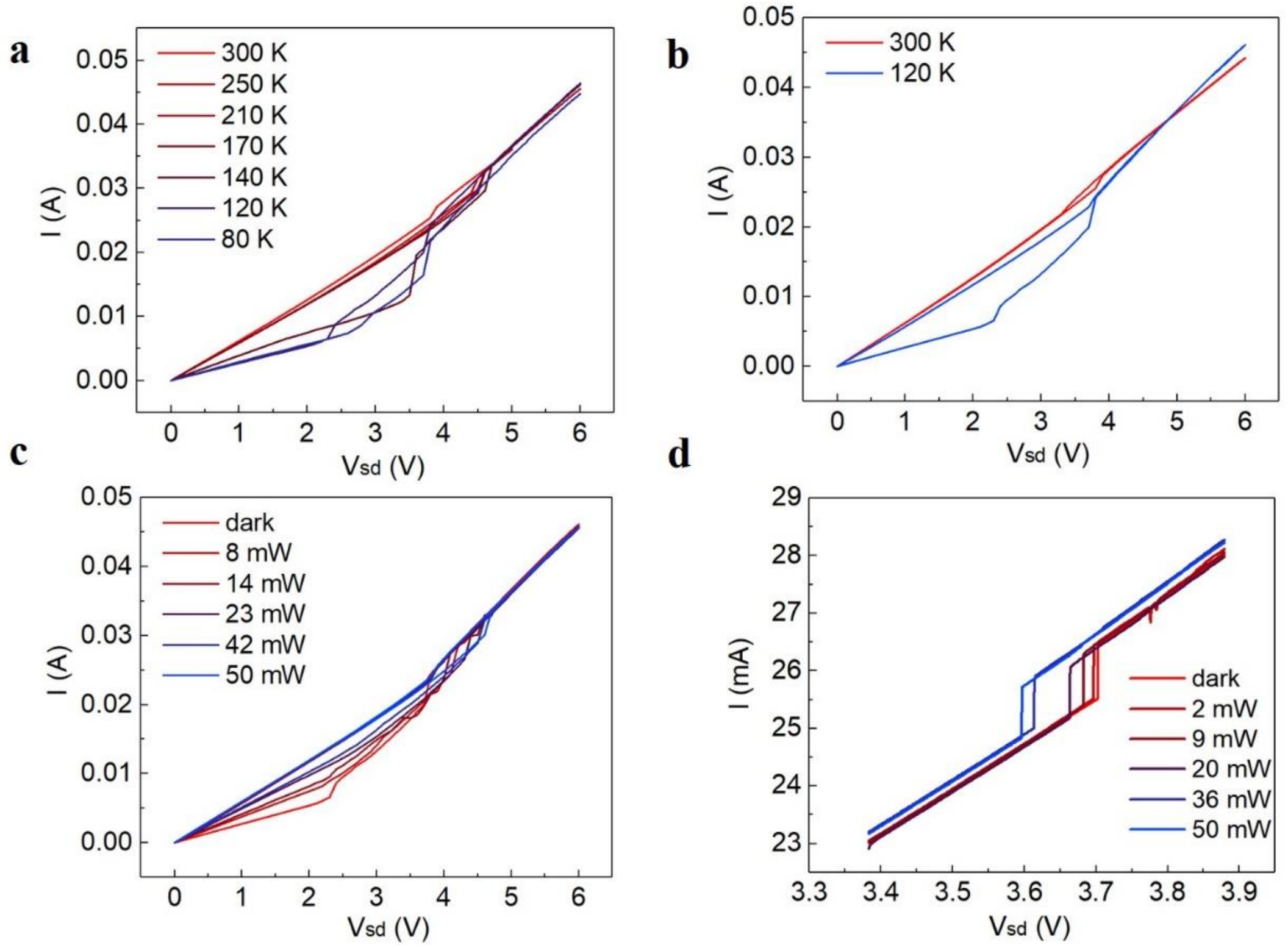

**Figure S15.** Photo reshaped CDW transition of 1T-$TaS_2$ with the thickness of 50 nm. (a) and (b) Temperature-dependent *I-V* curves. (c) Photo reshaped CDW phenomenon at 120 K. (d) Photo reshaped CDW phenomenon at room temperature.

15. Photo Response in (TaSe4)2I Devices at Different Working Region

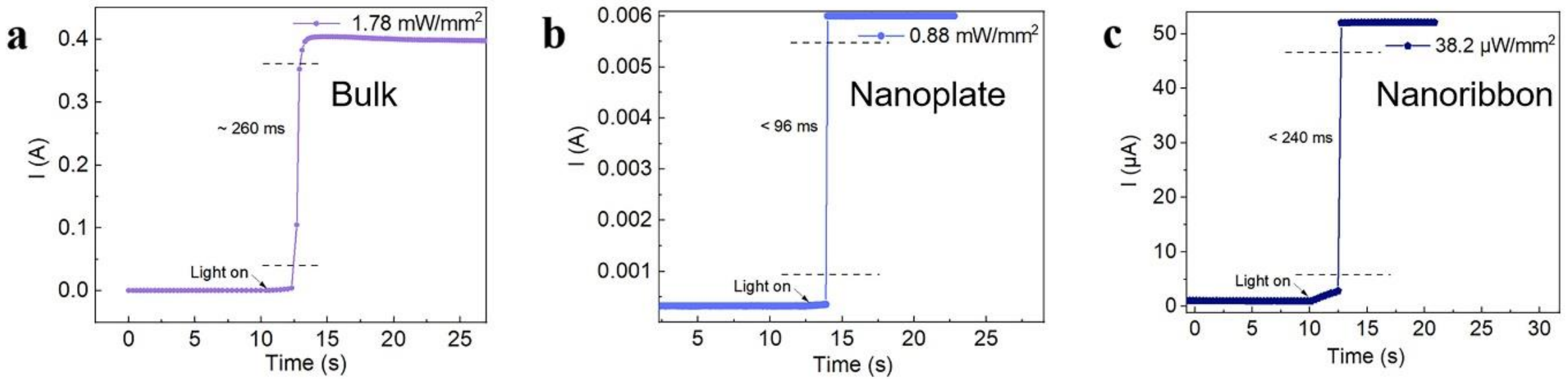

Figure S16. Photo response in ESCJ region of (TaSe4)2I bulk (refer to S2), nanoplate (refer to S5) and nanoribbon (refer to S6) at 120 K (a current limit of 6 mA is set to avoid sample breakdown) and 17 K respectively. The bias voltage is 3V (bulk), 3V (nanoplate) and 10 V (nanoribbon) respectively. The device channel area is 0.19 mm2 (bulk), 6 μm2 (nanoplate, ~ 80 nm-thick) and 3.9 μm2 (nanoribbon, ~ 35 nm-thick) respectively. The response time (in S5 and S6) is limited by accuracy of our source meter at present stage, actual response time should be smaller.

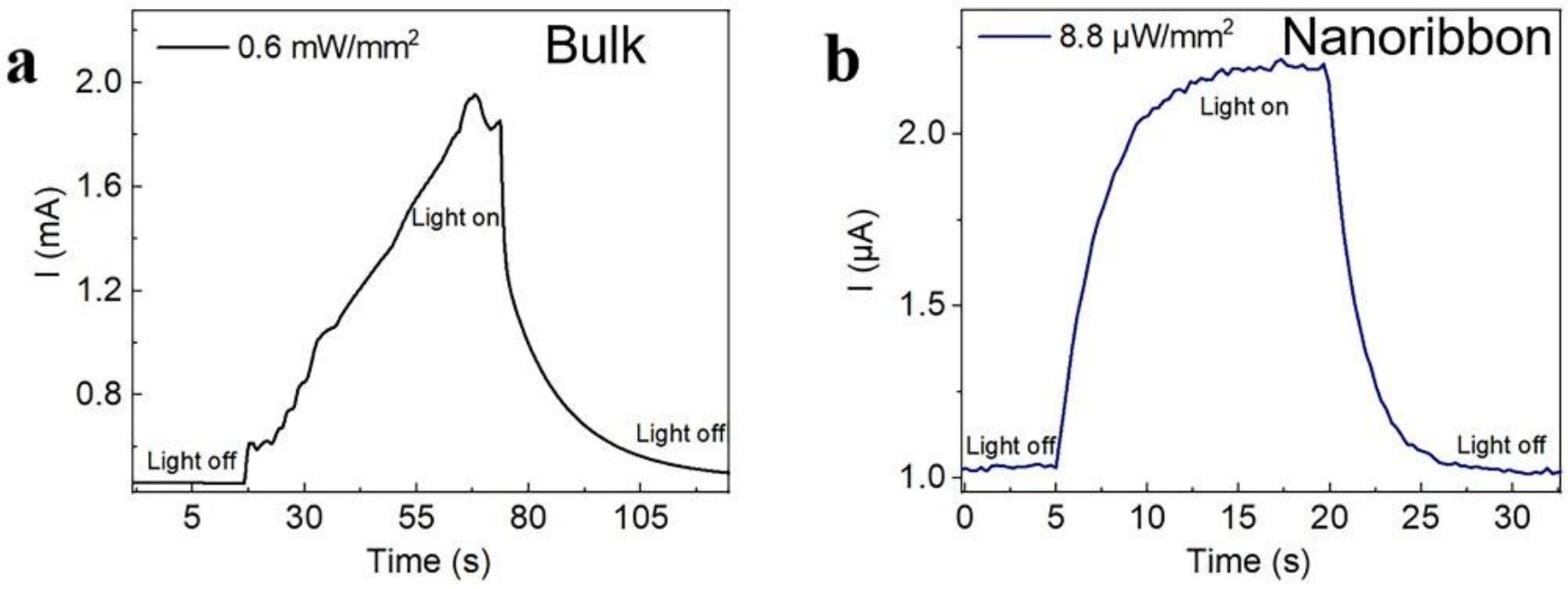

Figure S17. Photo response in CDW sliding region (before ESCJ) of (TaSe4)2I bulk (refer to S2) and nanoribbon (refer to S6) at 120 K and 17 K respectively. The bias voltage is 3V (bulk) and 10 V (nanoribbon) respectively. The device channel area is 0.19 mm2 (bulk) and 3.9 μm2 (nanoribbon, ~ 35 nm-thick) respectively.

Table S1. Parameters for low-dimensional based mid-infrared photodetectors.

| Material | Spectral range (μm) | Responsivity (A/W) | Specific Detectivity (Jones) | Response time (μs) | Light On/off ratio | Mechanism | Ref |
|---|---|---|---|---|---|---|---|
| HgCdTe (3D,77K) | 2-14 | 0.2-1.7@2-14 μm | $2\times10^8$-$3\times10^9$ @2-14 μm ($D^*_n$) | - | - | PV | [1] |
| TaAs (3D,RT) | 0.438-10.29 | $3\times10^{-4}$@3.1-10.6 μm | $1\times10^8$@3.1-10.6 μm ($D^*_d$) | $2\times10^5$ | - | PTE | [2] |
| $TaIrTe_4$ (2D,RT) | 0.638-10.6 | $3\times10^{-5}$@4 μm | $2.5\times10^6$@4 μm ($D^*_d$) | 26 | - | PTE | [3] |
| $PdSe_2$ (2D,RT) | 0.4-4.05 | $1.9\times10^{-3}$@4.05 μm | <$10^4$ @4.05 μm▲ ($D^*_n$) | $2.2\times10^5$ | 1.004▲ | PG | [4] |
| Graphene/Si QD (2D,RT) | 0.375-3.9 | 0.22@3.9 μm | <$10^3$ @3.9 μm ($D^*_n$) | $9\times10^6$ | ~1▲ | PG | [5] |
| BP (2D,77K) | 3.4-7.7 | 0.19@3.4 μm | - | < $1\times10^{-3}$ | 1.02▲ | PC | [6] |
| b-AsP (2D,RT) | 3.4-7.7 | 0.016@5 μm | - | $1\times10^{-4}$ | 1.002▲ | PC | [7] |
| $Nb_2SiTe_4$ (2D,RT) | 3.1 | 0.66@3.1 μm | - | >$5\times10^6$ | 1.01▲ | PC | [8] |
| Te (2D) | 0.52-3 | 353@3 μm | $2.3\times10^{11}$@3 μm ($D^*_d$); $3\times10^9$ @3 μm ($D^*_n$) | 52 | 2.58▲ | PC | [9] |
| $BP/MoS_2$ (2D,RT) | 0.25-4.1 | 0.9@3.6 μm | $7\times10^9$@3.6 μm ($D^*_n$) | 10 | - | PV | [10] |
| Graphene (2D,10K) | 0.532-10.3 | 0.4@10.3 μm | - | $3\times10^7$ | 1.23▲ | Impact ionization | [11] |
| BP/InSe (2D,10 K) | 4 | ~5 @4 μm▲ | - | 150 | $3\times10^4$▲ | Impact ionization | [12] |
| Bulk $TaS_2$ (2D,RT) | 0.532-118 | 1-3@1 μm-118 μm | - | $1.5\times10^{-3}$ @2.5μm | 1.56▲ | CDW | [13] |
| Thin $TaS_2$ (2D,RT) | 1-8 | ~0.5@1 μm-8 μm▲ | - | >$5\times10^6$ | - | CDW | [14] |
| InAs (1D,77K) | 0.83-3.13 | 0.6@3.113 μm | ~$10^{10}$@3.113 μm ($D^*_d$) | 80 | - | PC | [15] |
| InAsSb (1D,77K) | 1-3.5 | 0.19@2.39 μm | - | - | - | PC | [16] |
| Bulk $(TaSe_4)_2I$ (1D,120 K, S2) | 0.532-14 | a) $1.18\times10^3$ @4.64 μm (ESCJ region)<br>b) 11.93 @4.64 μm (Sliding region) | a) -<br>b) $2.92\times10^{10}$ @4.64 μm ($D^*_d$); ▲$10^8$ @4.64 μm ($D^*_n$) | a) $2.6\times10^5$ @4.64 μm<br>b) $4.2\times10^7$ @4.64 μm | 80 | CDW | This work |
| $(TaSe_4)_2I$ nanoplate (1D,120 K, S4) | 0.532-14 | 61.62@4.64 μm (Sliding region) | $1.13\times10^{10}$ @4.64 μm ($D^*_d$); ▲$10^8$ @4.64 μm ($D^*_n$) | $6.3\times10^6$ @4.64 μm | 1.51 | CDW | This work |
| $(TaSe_4)_2I$ nanoplate (1D,120 K, S5) | 0.532-14 | $1.1\times10^6$ @4.64 μm (ESCJ region) | - | <$9.6\times10^4$ @4.64 μm | 18.24 | CDW | This work |
| $(TaSe_4)_2I$ nanoribbon (1D,17 K, S6) | 0.532-14 | a) $3.4\times10^5$ @4.64 μm (ESCJ region)<br>b) $3.3\times10^4$ @4.64 μm (Sliding region) | a) -<br>b) $1.13\times10^{13}$ @4.64 μm ($D^*_d$); ▲$10^{11}$ @4.64 μm ($D^*_n$) | a) <$2.4\times10^5$ @4.64 μm<br>b) $5\times10^6$ @4.64 μm | 51.27 | CDW | This work |

Table notes:

1. RT: Room temperature, PV: Photovoltaic, PC: Photoconductive, PTE: Photo-thermoelectric, PG: Photogating.

2. Symbol ▲ indicates the estimation value form the data given in the corresponding reference.

3. The parameters here for $(TaSe_4)_2I$ devices here is given for a typical value, it can be optimized by increasing the bias voltage closer to $V_T$*.

4. The specific detectivity under bias voltage is often calculated by the formula $D^* = R\sqrt{A}/\sqrt{2eI_{dark}}$, as often used in low-dimensional materials[15, 17], to present the ability of detecting weak light signal. The detectivity calculated from noise current spectrum (referred as $D^*_n$) is often about 1-2 magnitude lower than the detectivity calculated by dark current (referred as $D^*_d$)[9, 18]. To make a fair comparison, we specially labeled the detectivity in Table S1. The symbol ▲ indicates the detectivity value ($D^*_n$) of our devices estimated two orders lower than $D^*_d$.

5. Since in sharp CDW transition region for CDW materials, the laser power triggers the sharp transition above a threshold power, it is improper to estimate the specific detectivity by using above formula, therefore detectivity in this region is not presented in Table S1.

6. The gain definition of avalanche photodetector (APD) is different in traditional photodetector case, in traditional photodetector case, the gain is calculated by using $\tau/t_r$ ($\tau$ is the time for the carrier trapping, and $t_r$ is the time for the transfer of carriers through the device channel). In APD case, the gain (multiplication factor) is defined by $M=(I_{ph} - I_{dark})/(I_{ph(0V)} - I_{dark(0V)})$ where $I_{ph}$ is the photocurrent, $I_{dark}$ is the dark current. For the avalache-like ESCJ region, to make a fair comparsion, we shoud compared our device with low-dimensinal material's APD. The detailed comparsion is shown in Table S2. The MIR region is red marked.

Table S2. Parameters comparsion for low-dimensional based APDs.

| **Material** | **Wavelength (μm)** | **Responsivity (A/W)** | **Gain** | **Working temperature** | **Ref** |
|---|---|---|---|---|---|
| InSe | 0.543 | 4.86 | 152 | 295 K | [19] |
| BP | 0.532 | 1.16 | 272 | 295 K | [20] |
| BP | 0.520 | 2 | 7 | 295 K | [21] |
| BP/InSe | 4 | 5▲ | $10^4$ | 10 K | [12] |
| InP/InAsP QD | 0.532 | 75▲ | $2.3 \times 10^4$ | 40 K | [22] |
| InP | 0.7 | 0.1▲ | $10^5$ | 300 K | [23] |
| Bulk $(TaSe_4)_2I$ (S2) | 4.64 | $1.18 \times 10^3$ | $6.3 \times 10^5$ | 120 K | This work |
| $(TaSe_4)_2I$ nanoribbon (S6) | 4.64 | $1.3 \times 10^6$ | $1.0 \times 10^6$ | 17 K | This work |

## Reference

[1] E. Theocharous, J. Ishii, N. Fox, *Infrared Physics & Technology* **2005**, 46, 309.

[2] S. Chi, Z. Li, Y. Xie, Y. Zhao, Z. Wang, L. Li, H. Yu, G. Wang, H. Weng, H. Zhang, J. Wang, *Adv. Mater.* **2018**, 30, 1801372.
[3] J. Lai, Y. Liu, J. Ma, X. Zhuo, Y. Peng, W. Lu, Z. Liu, J. Chen, D. Sun, *ACS Nano* **2018**, 4055.
[4] Q. Liang, Q. Wang, Q. Zhang, J. Wei, S. X. Lim, R. Zhu, J. Hu, W. Wei, C. Lee, C. Sow, W. Zhang, A. T. S. Wee, *Adv. Mater.* **2019**, 31, e1807609.
[5] Z. Ni, L. Ma, S. Du, Y. Xu, M. Yuan, H. Fang, Z. Wang, M. Xu, D. Li, J. Yang, W. Hu, X. Pi, D. Yang, *ACS Nano* **2017**, 11, 9854.
[6] X. Chen, X. Lu, B. Deng, O. Sinai, Y. Shao, C. Li, S. Yuan, V. Tran, K. Watanabe, T. Taniguchi, D. Naveh, L. Yang, F. Xia, *Nat Commun* **2017**, 8, 1672.
[7] S. Yuan, C. Shen, B. Deng, X. Chen, Q. Guo, Y. Ma, A. Abbas, B. Liu, R. Haiges, C. Ott, T. Nilges, K. Watanabe, T. Taniguchi, O. Sinai, D. Naveh, C. Zhou, F. Xia, *Nano Lett.* **2018**, 18, 3172.
[8] M. Zhao, W. Xia, Y. Wang, M. Luo, Z. Tian, Y. Guo, W. Hu, J. Xue, *ACS Nano* **2019**, 13, 10705.
[9] L. Tong, X. Huang, P. Wang, L. Ye, M. Peng, L. An, Q. Sun, Y. Zhang, G. Yang, Z. Li, F. Zhong, F. Wang, Y. Wang, M. Motlag, W. Wu, G. J. Cheng, W. Hu, *Nat Commun* **2020**, 11, 2308.
[10] J. Bullock, M. Amani, J. Cho, Y.-Z. Chen, G. H. Ahn, V. Adinolfi, V. R. Shrestha, Y. Gao, K. B. Crozier, Y.-L. Chueh, A. Javey, *Nat. Photonics* **2018**, 12, 601.
[11] B. Y. Zhang, T. Liu, B. Meng, X. Li, G. Liang, X. Hu, Q. J. Wang, *Nat Commun* **2013**, 4, 1811.
[12] A. Gao, J. Lai, Y. Wang, Z. Zhu, J. Zeng, G. Yu, N. Wang, W. Chen, T. Cao, W. Hu, D. Sun, X. Chen, F. Miao, Y. Shi, X. Wang, *Nat. Nanotechnol.* **2019**, 14, 217.
[13] D. Wu, Y. Ma, Y. Niu, Q. Liu, T. Dong, S. Zhang, J. Niu, H. Zhou, J. Wei, Y. Wang, Z. Zhao, N. Wang, *Sci Adv* **2018**, 4, eaao3057.
[14] X. Wang, H. Liu, J. Wu, J. Lin, W. He, H. Wang, X. Shi, K. Suenaga, L. Xie, *Adv. Mater.* **2018**, 30, 1800074.
[15] H. Fang, W. Hu, P. Wang, N. Guo, W. Luo, D. Zheng, F. Gong, M. Luo, H. Tian, X. Zhang, C. Luo, X. Wu, P. Chen, L. Liao, A. Pan, X. Chen, W. Lu, *Nano Lett.* **2016**, 16, 6416.
[16] W.-J. Lee, P. Senanayake, A. C. Farrell, A. Lin, C.-H. Hung, D. L. Huffaker, *Nano Lett.* **2016**, 16, 199.
[17] X. Zhang, H. Huang, X. Yao, Z. Li, C. Zhou, X. Zhang, P. Chen, L. Fu, X. Zhou, J. Wang, W. Hu, W. Lu, J. Zou, H. H. Tan, C. Jagadish, *ACS Nano* **2019**, 13, 3492.
[18] L. Liang, X. Niu, X. Zhang, Z. Wang, J. Wu, J. Luo, *Adv. Opt. Mater.* **2022**, 10, 2201342.
[19] S. Lei, F. Wen, L. Ge, S. Najmaei, A. George, Y. Gong, W. Gao, Z. Jin, B. Li, J. Lou, J. Kono, R. Vajtai, P. Ajayan, N. J. Halas, *Nano Lett.* **2015**, 15, 3048.
[20] M. R. M. Atalla, S. J. Koester, presented at *2017 75th Annual Device Research Conference (DRC)*, 25-28 June 2017, **2017**.
[21] J. Jia, J. Jeon, J.-H. Park, B. H. Lee, E. Hwang, S. Lee, *Small* **2019**, 15, 1805352.
[22] G. Bulgarini, M. E. Reimer, M. Hocevar, E. P. A. M. Bakkers, L. P. Kouwenhoven, V. Zwiller, *Nat. Photonics* **2012**, 6, 455.
[23] S. J. Gibson, B. van Kasteren, B. Tekcan, Y. Cui, D. van Dam, J. E. M. Haverkort, E. P. A. M. Bakkers, M. E. Reimer, *Nat. Nanotechnol.* **2019**, 14, 473.